\documentclass[fleqn,usenatbib]{mnras}

\usepackage{newtxtext,newtxmath}

\usepackage[T1]{fontenc}
\usepackage{ae,aecompl}

\usepackage{graphicx}  
\usepackage{dcolumn}   
\usepackage{bm}        
\usepackage{comment}
\usepackage{amssymb}   
\usepackage{amsmath}
\usepackage{tabularx}
\usepackage{tabulary}
\usepackage[usenames,dvipsnames]{color}
\usepackage[normalem]{ulem}
\usepackage{lastpage}
\newcommand{\msun}{{\rm M}_{\odot}}

\newcommand{\be}{\begin{equation}}
\newcommand{\ee}{\end{equation}}
\newcommand{\bi}{\begin{list}{\labelitemi}{\leftmargin=1em}\setlength{\itemsep}{-3pt}}
\newcommand{\ei}{\end{list}}

\newcommand{\beqn}{\begin{eqnarray}}
\newcommand{\eeqn}{\end{eqnarray}}

\newcommand{\trh}{t_{\rm rh}}
\newcommand{\trhn}{t_{\rm rh,0}}
\newcommand{\rh}{r_{\rm h}}

\newcommand{\rhdot}{\dot{r}_{\rm h}}
\newcommand{\rhoh}{\rho_{\rm h}}
\newcommand{\mst}{m_\star}
\newcommand{\mmeanst}{\langle\mst\rangle}
\newcommand{\mall}{m_{\rm all}}
\newcommand{\mmean}{\langle \mall\rangle}
\newcommand{\vesc}{v_{\rm esc}}

\newcommand{\kms}{{\rm km\,s}^{-1}}
\newcommand{\myr}{{\rm Myr}}
\newcommand{\pc}{{\rm pc}}

\def\dr{{\rm d}}

\def\emacss{{\sc emacss}}
\def\python{{\sc python}}
\def\fortran{{\sc fortran}}
\def\nbsix{{\sc nbody6}}
\def\emcee{{\sc emcee}}
\def\model{{\sc clusterBH}}
\def\modelc{{\sc  BHBdynamics}}
\def\modelb{{\sc clusterBHBdynamics}}
\def\CB{C_{\rm B}}
\def\fbh{f_{\rm BH}}
\def\feh{{\rm [Fe/H]}}
\def\fretm{f_{\rm ret}^M}
\def\fret{f_{\rm ret}^N}
\def\Mcl{M_{\rm cl}}    
\def\Ncl{N_{\rm cl}}    
\def\Mcldot{\dot{M}_{\rm cl}}    
\def\mbh{m}    
\def\mns{m_{\rm NS}}    
\def\mb{m_{\rm b}}    
\def\Mbh{M_{\rm BH}}   
\def\Mbhn{M_{\rm BH,0}}   
\def\Mbhdot{\dot{M}_{\rm BH}}   
\def\Mst{M_\star}   
\def\Mstdotsev{\dot{M}_{\star, \rm sev}}   
\def\mlo{m_{\rm lo}}    
\def\mup{m_{\rm up}}    
\def\qul{q_{\rm ul}}
\def\qub{q_{\rm ub}}
\def\qml{q_{\rm ml}}
\def\qmb{q_{\rm mb}}

\def\qlb{q_{\rm lb}}
\def\mmax{m_{\rm max}}    
\def\rhdot{\dot{r}_{\rm h}}
\def\rhdotsev{\dot{r}_{\rm h, sev}}
\def\rhdotrlx{\dot{r}_{\rm h, rlx}}
\def\sigmak{\sigma_{\rm k}}
\def\sigmans{\sigma_{\rm NS}}
\def\vkick{v_{\rm kick}}
\def\tcc{t_{\rm cc}}
\def\tsev{t_{\rm sev}}

\def\hyp#1#2#3#4{{_2}F_1\left(#1, #2; #3; #4\right)}
\def\ntrh{N_{\rm rh}}

\def\mgq#1{{\it\textcolor{magenta}{MG:#1}}}

\title[]{Population synthesis of black hole binary mergers from  star clusters}

\author[Antonini and Gieles]{
Fabio Antonini$^{1,2}$ and 
Mark Gieles$^{2,3,4}$
\\
$^1$School of Physics and Astronomy, Cardiff University, Cardiff, CF24 3AA, United Kingdom
\\
$^2$Faculty of Engineering and Physical Sciences, University of Surrey, Guildford, Surrey, GU2 7XH,
United Kingdom \\
$^3$Institut de Ci\`{e}ncies del Cosmos (ICCUB), Universitat de Barcelona, Mart\'{i} i Franqu\`{e}s 1, 08028 Barcelona, Spain\\
$^4$ICREA, Pg. Lluis Companys 23, 08010 Barcelona, Spain.
}

\date{Accepted XXX. Received YYY; in original form ZZZ}

\pubyear{2018}

\begin{document}
\label{firstpage}
\pagerange{\pageref{firstpage}--\pageref{lastpage}}
\maketitle

\begin{abstract}
Black hole (BH) binary mergers  formed through dynamical interactions in dense star clusters
are believed to be one of the main sources 
of gravitational waves  for Advanced LIGO and Virgo.
Here we present a fast numerical  method  for simulating the  evolution 
of  star clusters with  BHs, including a model for the dynamical formation and merger of BH binaries. 
Our method is based on  H\'{e}non's principle of balanced evolution, 
according to which the flow of energy within a cluster must 
be balanced by the energy production inside its core. 
Because  the heat  production in the core is powered 
by the BHs, one can then  link  the evolution of the cluster to the evolution of its BH population.
This allows us to construct evolutionary tracks of the cluster properties including its BH population and its effect on the cluster and, at the same time, determine the merger rate of  BH binaries 
as well as their eccentricity distributions. The model is publicly available
and includes the effects of a BH mass spectrum, mass-loss due to stellar evolution, the ejection of BHs
due to natal and dynamical  kicks, and 
relativistic corrections during binary-single encounters.
We validate our method using direct  $N$-body simulations, and  find it to be  in excellent  agreement with results from recent 
Monte Carlo models of globular clusters. 
This establishes our new method as a robust tool for the  study of  BH dynamics 
in star clusters and the  modelling of gravitational wave sources  produced
  in these systems. Finally, we compute the rate
and eccentricity distributions of merging BH binaries for a wide range
of cluster initial conditions, spanning more than
two orders of magnitude in mass and radius.
\end{abstract}              



 \section{Introduction}
The advanced gravitational-wave observatories  LIGO and Virgo are routinely  detecting 
gravitational waves (GWs) from the merger of   black hole (BH) binaries \citep{2010CQGra..27q3001A,2015CQGra..32b4001A,2016PhRvL.116f1102A,2016PhRvX...6d1015A}. The planned detector 
 KAGRA will soon become operative \citep{2018LRR....21....3A} and the 
Laser Interferometer Space Antenna (LISA), planned to launch in  2030s, will detect GWs in space \citep{2017arXiv170200786A}, allowing to observe the BH mergers at lower
 frequencies and opening the doors of multi-band GW astrophysics   \citep{2016PhRvL.116w1102S}.
 
 The detection of GWs has  opened new perspectives for the study of compact object binaries, and has generated great interest in understanding how these sources form.
 A number of formation scenarios have been proposed for the formation of BH binary mergers.
 These include: the evolution of massive star binaries in the field of a galaxy \citep{Belczynski2010,Dominik2012,Marchant2016,Mandel2016a,2018PhRvD..98h4036G,2019arXiv190201864M}, the evolution
 of hierarchical multiple field stars \citep{Silsbee2017,2017ApJ...841...77A,2019MNRAS.483.4060L,2019MNRAS.486.4781F},  and dynamical few-body interactions in the  core of  a dense 
 star cluster such as young star clusters  \citep{2014MNRAS.441.3703Z,Kimpson2016,2017MNRAS.467..524B,2018MNRAS.473..909B,2018MNRAS.481.5123B,2019MNRAS.487.2947D}, 
 nuclear star clusters \citep{2009ApJ...692..917M,2016ApJ...831..187A,2018MNRAS.474.5672L},
 and globular clusters \citep[GCs;][]{Kulkarni1993,Sigurdsson1993,PortegiesZwart2000,Downing2011,Rodriguez2015a,2017MNRAS.464L..36A}.
 In this paper we are concerned with  the latter scenario, i.e. binary BH formation in star clusters of all masses.
 
 In the dynamical formation scenario, the BH binaries are assembled through 
 three body  processes in the core of a star cluster, and subsequently 
harden and merge via 
 dynamical binary-single  interactions.
Recent studies suggest that such binaries might account for many, or perhaps even most, binary BH mergers
so far detected by LIGO-Virgo \citep{2018PhRvL.121p1103F,2018ApJ...866L...5R,2019ApJ...873..100C}. 
Moreover, a fraction of BH binary mergers from clusters are expected to have a a finite eccentricity when they first enter the frequency band of current detectors. 
 It has been argued therefore  that 
GW observations of eccentric binary BH mergers will provide evidence for a dynamical formation of these systems  \citep{2014ApJ...781...45A,2016PhRvD..94f4020N,2016ApJ...830L..18B,2018PhRvD..97j3014S,2018PhRvL.120o1101R}.
Major effort around the detection and characterisation of eccentric BH binaries
is currently
 underway \citep{2017PhRvD..96d4028C,2018PhRvD..98d4015H,2019arXiv190107038H}.
 
Theoretical predictions for the dynamical formation channel are currently based
either on   Monte Carlo  or direct $N$-body simulations of the long-term 
evolution of star clusters \citep{Giersz1998,2000ApJ...540..969J,Aarseth2012,Giersz2013}. These methods
allow an accurate treatment of stellar dynamical relaxation and strong encounters
and  have therefore the advantage  that they can solve the long-term evolution of a star cluster self-consistently. Moreover,
 they include recipes for the effect of stellar and binary evolution, Galactic tides, 
primordial binaries, and relativistic corrections during strong encounters. The drawback is that the time requirements for a simulation of a realistic set of cluster models are  prohibitive.
For this reason, current predictions for the merger rate and source properties are  based on a limited  set of cluster initial conditions which 
do not allow to explore the relevant parameter space and uncertainties
on the cluster initial conditions \citep[e.g.,][]{2018PhRvL.121p1103F,2018ApJ...866L...5R}.
For example, \citet{2018PhRvL.121p1103F}
ignored
the dependence of the binary merger rate on the cluster initial
radius and its evolution.
\citet{2018ApJ...866L...5R} derived their binary BH merger rate 
assuming that $50\%$ of clusters form with a virial  radius of
$r_{\rm v}=1\,$pc and the rest with $2\,$pc. To fully explore the parameter space that controls dynamically formed BH mergers, a significantly faster recipe is required, without giving in too much on the accuracy of the obtained results.

In this paper we present a new, fast computational method for
the evolution of a star cluster with BHs, including a prescription
for the dynamical evolution of the BH binaries.
Any of our model takes less than a second to complete (using 
a commercial laptop). In comparison,
the typical wall-clock computation time for a
full $N$-body model of $\sim 10^6~\msun$ cluster is about a year
on a super-computer with Graphical Processing Units  \citep[GPUs,][]{Wang2016}, and
a similar Monte Carlo cluster model {  will  require about one week on a standard multi-core PC}.

Our method relies on  H\'{e}non's principle \citep{1975IAUS...69..133H}, which states that {  after an initial settling phase}
the rate of heat generation in the core 
is a constant fraction of the total cluster energy per half-mass relaxation time.
\citet{2013MNRAS.432.2779B} showed that in {  this so-called} `balanced evolution' {  phase} the heat is produced
by the BHs, and one can therefore link the  evolution of the BH population
 to the properties of the
cluster itself.  
This allows us to construct a series of coupled first-order differential equations to express the time evolution of a cluster mass, radius, and the total mass in BHs.
We further assume that during binary-single interactions
the   eccentricity of the BH binaries follows that  
of a so-called thermal distribution $N(e)\propto e$ \citep[e.g.,][]{Heggie1975}. 
Based on this  assumption, we  derive 
 the  merger rate, and the eccentricity distribution of the BH binaries and their relation
 to a cluster global properties.

The paper is organised as follows. 
In Section~\ref{clev} we present the model for the co-evolution of a BH population and its host cluster,
and use direct $N$-body simulations to validate
this simple model. 
 Section~\ref{analy}  describes the analytical prescriptions
 to determine the  rate  and eccentricity distributions 
of  BH binary mergers formed via binary-single interactions. 
In Section~\ref{MCcompS} we compare the results
of our model to those of independent Monte Carlo simulations from published literature.
Finally, in Section~\ref{AI} we 
make predictions for the merger rate and eccentricities of BH binaries for a wide range 
of cluster initial conditions, and discuss how these distributions
are linked to a cluster properties.

\section{Evolution of star clusters with a black hole population} \label{clev} 
\subsection{Philosophy of the model}
In this section we present a fast model for the co-evolution of a BH population and its host cluster.
To achieve speed, we only evolve several bulk properties of the cluster and not its internal structure.  We limit the model to the cluster properties that are most relevant for the formation and evolution of binary BHs. 
The hard-soft boundary of binaries is set by the velocity dispersion of the cluster, which is proportional to $\sqrt{\Mcl/\rh}$, where $\Mcl$ is the total cluster mass and $\rh $ its half-mass radius. The binding energy, and therefore, the semi-major axis of escaping BH binaries depends on the central escape velocity ($\vesc$) of the cluster (see \citealt{Rodriguez2016a}), which is also proportional to $\sqrt{\Mcl/\rh}$. The condition for in-cluster BH binary mergers to be efficient can also be expressed in $\Mcl$ and $\rh$ (see \citealt{2019MNRAS.486.5008A}). We therefore model the evolution of $\Mcl$ and $\rh$. We will solve differential equations for $\Mcldot$ and $\rhdot$, as is done in  the fast cluster evolution model \emacss\ \citep{Alexander2012, Gieles2014, Alexander2014}. Here we  add the evolution of the BH population, and its effect on the  evolution of the cluster. We approximate the cluster by a two component system: a light component, consisting of the stars, white dwarfs and neutron stars, and a heavy component of BHs.

\citet{2013MNRAS.432.2779B} showed that once a star cluster achieves the  balanced evolution\footnote{In their models this is the moment the BH core collapses, while in clusters without BHs this is the moment of the collapse of the visible (i.e. stellar) core.} phase, the total mass of the BH population in a cluster ($\Mbh$) depends on $\Mcl$ and the half-mass relaxation time-scale ($\trh$)  of the cluster as $\Mbhdot \propto \Mcl/\trh$. This dependence of $\Mbh$ on the  cluster properties, rather than the properties of the core where the BH binaries reside, is because the heat production in the core is set by the properties of a cluster as a whole \citep{1961AnAp...24..369H}, and the BH binaries provide the heat via dynamical interactions with other BHs, resulting in ejections of BHs and binaries (more on this in Section~\ref{ssec:coevolv}).  This discovery by Breen \& Heggie allows us to relate the evolution of the BH population to the cluster properties and we therefore jointly solve  for $\Mcl$, $\rh$ and $\Mbh$ from the expressions for their time derivatives. 

We restrict ourselves in this paper to isolated star clusters, avoiding the complication of the Galactic tidal field. This will result in unrealistic cluster properties at redshift $z\simeq0$ for clusters that are strongly tidally limited, but most of the BH binaries that are relevant for GW detections are produced at high $z$, when the clusters are dense compared to the tidal density \citep{2010MNRAS.408L..16G}, and then GC evolution is comparable to that of isolated GCs \citep*{2011MNRAS.413.2509G}. Clusters with low densities are more affected by the tides, but less relevant for GWs.
In addition, the most massive GCs, which are most important for GW production \citep{2016ApJ...831..187A}  are still largely unaffected by the tidal field at $z=0$ \citep{2011MNRAS.413.2509G}, providing further support for our assumption. 

We first present in Section~\ref{ssec:fretm}  a prescription for the fraction of the initial $\Mbh$ that is retained after natal kicks and in Section~\ref{ssec:coevolv} we discuss the co-evolution of $M$, $\rh$ and $\Mbh$. In Section~\ref{ssec:nbody} we compare the model to a series of direct $N$-body simulations and use these to determine a few model parameters.

 \subsection{Retention after supernova kicks} \label{ssec:fretm}
For the initial conditions we need $M$, $\rh$ and $\Mbh$ at $t=0$, i.e. when the cluster forms. The value of $\Mbh$ is zero initially, because all stars are on the main sequence, but for simplicity we assume that all BHs are already in place at formation avoiding the need for describing BH formation  in the first $\sim20~$Myr. The initial value of $\Mbh$ then depends on  the stellar initial mass function (IMF) and the initial-final mass relation (IFMR), which depends on metallicity (\feh). At lower \feh, BHs  are more massive \citep[e.g.][]{Spera2015a} such that -- for a given IMF -- $\Mbh$ is  larger.  The \feh\ and IMF dependence can be parameterised. {  Here, we focus on a canonical \citet{Kroupa2001} IMF, for which a fraction $f_0 \simeq0.06$ of the initial $\Mcl$ ends up in BHs for metal-poor GCs.}

Because BHs receive natal kicks, and the condition for escape depends on $\vesc$,  we  need a relation for the retention of the  mass  retention fraction after supernova (SN) kicks ($\fretm$)\footnote{The term `retention fraction' is often used in references to the number fraction, hence we use the super-script $M$ to make it clear that we refer to a mass fraction.}. Although the magnitude of BH natal kick velocity  ($\vkick$) is not known, there is some consensus that BH kicks are smaller than those of neutron stars 
\citep{2016MNRAS.456..578M}, and it likely depends on the mass of the BH ($\mbh$), with the more massive BHs receiving smaller kicks. If the escape velocity from the centre of the cluster ($\vesc$), where most BHs are expected to form, is much larger than  $\vkick$ of the lowest mass BH, then $\fretm\simeq1$. 
 If $\vesc$ is much smaller than $\vkick$ of the most massive BHs, then $\fretm\simeq0$. 
In the GC-mass range, $\vesc$ is likely to be in the intermediate regime, and for a given BH mass $\mbh$, there is then a distribution of $\vkick$ values. Both of these effects need to be captured by our expression for $\fretm(\vesc)$. 

To proceed, we assume that $\vkick$ is drawn from a Maxwell-Boltzmann distribution, $P_{\rm B}(\vkick\vert\sigmak)$, with dispersion $\sigmak$. We then assume that BHs receive the same momentum kick as neutron stars \citep{Fryer2001}, such that $\sigmak(\mbh)=\sigmans\times\mns/\mbh$, for which we adopt $\sigmans=265\,\kms$ is the dispersion for neutron star kicks \citep{Hobbs2005} and $\mns=1.4~\msun$ is the mass of a neutron star. The resulting kicks are smaller than in the fallback scenario \citep{Dominik2013}, but for our simple model the assumption of a constant momentum kick is preferred, because it allows us to derive a simple expression for $\fretm(\vesc)$, without detailed knowledge of the fallback fraction of BHs with different masses. We ignore direct collapse SNe in this version of the model, but note that more detailed kick prescriptions will be considered in future versions of the model. 

The retained number fraction of BHs with mass $\mbh$ is then given by the integral over $P_{\rm B}$ from zero to  $\vesc$
\begin{equation}
\fret(\vesc, \mbh) = \int_0^{\vesc} P_{\rm B}(v^\prime\vert\sigmak)\dr v^\prime = \CB(\vesc\vert\sigmak), 
\end{equation}
where $\CB(\vesc\vert\sigmak)$ is the cumulative distribution function of the Maxwell-Boltzmann distribution. 

To find $\fretm$, we define the BH mass function after SN kicks, $\phi(\mbh)$, defined as the number of BHs in an interval $(\mbh, \mbh+\dr\mbh)$. This mass function follows from the birth BH mass function, $\phi_0(\mbh)$, as

\begin{align}
\phi(\vesc,\mbh) &= \CB(\vesc\vert\sigmak)\phi_0(\mbh),\label{eq:phi}\\
                       &\simeq \frac{\phi_0(\mbh)}{(\mb/\mbh)^3 + 1}.\label{eq:phiappr}
\end{align}
In the last step we used an approximation, where $\mb = (9\pi/2)^{1/6}\sigmans\mns/\vesc$ is the mass below which the BH mass function is affected by kicks. {  For all $\phi_0$, this approximation underestimates the exact result by only $\sim10\%$ at $m\simeq\mb$, and quickly approaches to the exact result for other values of $m$. } This simple approximation for $\phi$ serves to derive the maximum BH mass from $\Mbh$ when we include escape of BHs.

\begin{figure}
\includegraphics[width=8cm]{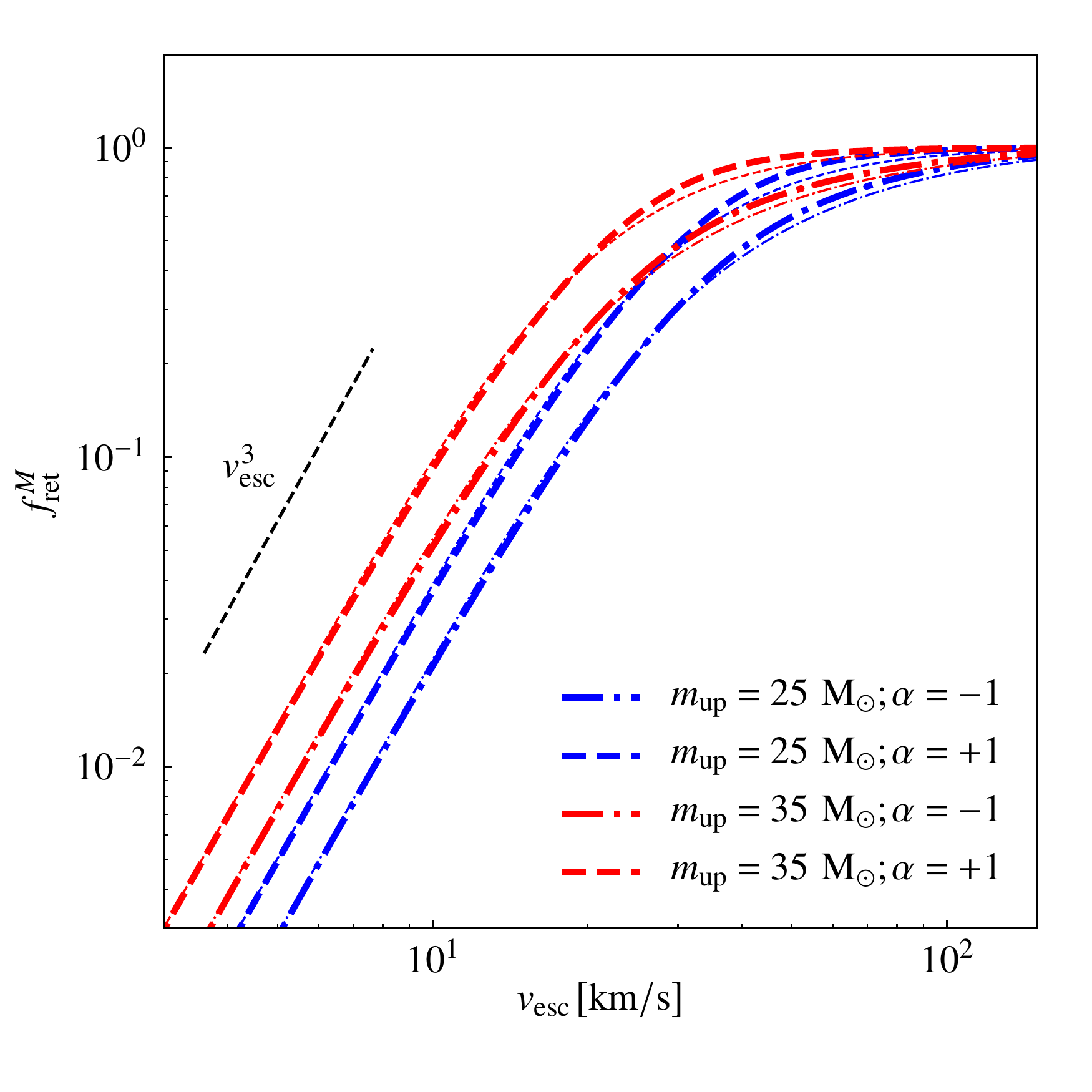}
\caption{Fraction of BH mass retained after SNe as a function of $\vesc$. The thick (blue and red) lines show results from {  equation~(\ref{eq:phi})}, for different $\mup$ and BH mass function slope $\alpha$. The thin lines show the approximation from equation~(\ref{eq:fretmbhapprox}).}
\label{fig:fretm}
\end{figure}

 {  The mass retention fraction $\fretm$ is then found from  integrating $\phi$ over all BH masses,
 \begin{align}
&\fretm(\vesc)=\int_{\mlo}^{\mup}\phi(\mbh)\mbh \dr \mbh \bigg/\int_{\mlo}^{\mup} \phi_0(\mbh) \mbh \dr \mbh\ .
\label{eq:fretmbh0}
\end{align}
 Assuming a power-law $\phi_0\propto m^\alpha$ and the approximation for $\phi$ from equation~(\ref{eq:phiappr}), then equation\ (\ref{eq:fretmbh0}) gives}
\begin{align}
\fretm(\vesc)&\simeq
\int_{\mlo}^{\mup} \frac{\mbh^{\alpha+1}}{(\mb/\mbh)^3 + 1}\dr \mbh\bigg/\int_{\mlo}^{\mup} \mbh^{\alpha+1} \dr \mbh\nonumber 
\\
&=\begin{cases}
\displaystyle\ln{\left[\left(\qub^3+1\right)/\left(\qlb^3+1\right)\right]}/\ln\left(\qul^3\right),&\alpha=-2,\vspace{0.2cm}\\
\displaystyle1 - \left[\qul^{\alpha+2}h(\qub) - h(\qlb)\right]/\left(\qul^{\alpha+2}-1\right), &\alpha\ne-2,
\end{cases} 
\label{eq:fretmbhapprox}
\end{align}
Here $\mlo$ and $\mup$ are the lower and upper limit of $\phi_0$, respectively, and {  $\qul=\mup/\mlo$, $\qub=\mup/\mb$, $\qlb=\mlo/\mb$ and $h(x) = \hyp{1}{(\alpha+2)/3}{(\alpha+5)/3}{-x^{3}}$}, with  $\hyp{a}{b}{c}{x}$ a hypergeometric function. {  This function diverges for negative integer values of $c$ and for $c=0$, hence the  result for $\fretm$ in equation~(\ref{eq:fretmbhapprox}) for $\alpha\ne-2$ also excludes the values $\alpha = -5, -8, -11, \dots$,  but because such steep mass functions are not expected for BHs, we refrain from giving the full solution for those values. } 
In Fig.~\ref{fig:fretm} we show results {  obtained from} numerical integrations of {  equation~(\ref{eq:phi})} in thick lines, and the approximation of equation~(\ref{eq:fretmbhapprox}) with thin lines. We assumed a power-law $\phi_0$, adopting both a declined and a rising BH mass function ($\alpha=-1$ and $\alpha=+1$, respectively), between $\mlo=3~\msun$ and $\mup$, for which we used $\mup=25~\msun$ and $\mup=35~\msun$.  From Fig.~\ref{fig:fretm} we see that at low $\vesc$, the retention fraction can be well approximated by $\fretm\propto \vesc^3$, which is the leading order term of $\CB(\vesc\vert\sigmak)$. More specifically we find 
\begin{equation}
\fretm(\vesc)\propto \frac{\vesc^3}{\sigmans^3}\frac{ \langle  \mbh^4 \rangle}{\langle \mbh \rangle}.
\label{eq:fretmapprox}
 \end{equation}
This  shows that at low $\vesc$, $\fretm$ depends on the cluster properties as $\vesc^3 \propto (M/\rh)^{3/2}$ and is sensitive to the most massive BH, because for top-heavy $\phi$ (i.e. $\alpha>-2$) and $\mup>>\mlo$ we find $\langle \mbh^4\rangle/\langle \mbh \rangle\propto \mup^3$. This sensitivity to $\mup$  implies that {  for a given $\vesc$,} $\fretm$ is higher in low-metallicity GCs, for which BHs are more massive. This statement is valid for a continuous $\phi$, since $\fretm$ can be low if -- for example -- a single, massive BH forms among a population of low-mass BHs.

\subsection{Co-evolution of the cluster and its BH population}
\label{ssec:coevolv}
In this section we present a model for the temporal evolution of $\Mcl$, $\rh$ and $\Mbh$. We consider clusters in isolation, which lose mass by stellar evolution and BHs via dynamical interactions. We define expressions for $\Mcldot$, $\rhdot$ and $\Mbhdot$, which are then integrated numerically. 

\subsubsection{Stellar mass loss}
We assume that the cluster  consists of two types of members: BHs and  {  all the other members} (i.e. other stellar remnants and stars). Each contribute a mass of $\Mbh$ and $\Mst$, respectively, such that $\Mcl=\Mst+\Mbh$.
We assume that as a result of stellar mass loss,  $\Mst$ evolves as 
a power of time, such that
 \begin{equation}
\Mstdotsev=
\begin{cases}
0, &t<\tsev,\\
\displaystyle -\nu {\Mst\over t}, &t\ge \tsev,
\end{cases}
\label{eq:Mstdotsev}
\end{equation}
with $\tsev\simeq 2\,\rm Myr$ and $\nu\simeq0.07$,  depending slightly on
\feh. We further assume
that {  as a result of stellar mass loss} the cluster expands adiabatically, such that 
 \begin{equation}
\rhdotsev=-{\Mstdotsev\over\Mcl}\rh.
\label{eq:rhdotsev}
\end{equation}
Note that we here assumed that stellar mass-loss occurs throughout the cluster, and the expansion rate could be higher if the cluster is mass segregated and mass loss occurs more central. This can be included \citep[see][]{Alexander2014}, but here we do not implement this, because  in the presence of BHs and for high density clusters, the evolution becomes quickly dominated by relaxation driven expansion, as we will show in the next section.

\subsubsection{Relaxation}
After several relaxation time-scales have elapsed, the core of the cluster starts producing energy in order to sustain the relaxation process. We define the start of this balanced evolution as 
\begin{equation}
\tcc = \ntrh\trhn,
\end{equation}
where $\ntrh$ is a constant of order unity that we will fit to $N$-body models in Section~\ref{ssec:nbody} and $\trhn$ is the initial half-mass relaxation time-scale.  It would be more accurate to express $\tcc$ in terms of the number of elapsed relaxation time-scales \citep{Alexander2014}, but for simplicity we here express $\tcc$ in the timescale that can be straightforwardly evaluated at the start of the model computation. 
The  start of balanced evolution corresponds to the collapse of the core, which is dominated by BHs {  in case $\fretm>0$}. This type of core collapse occurs well before the collapse of the core of visible stars, which coincides with the moment that all BHs are ejected \citep{2013MNRAS.432.2779B}. 

The half-mass relaxation time-scale is the average relaxation time-scale within $\rh$, which is given by \citep[][]{1971ApJ...164..399S}
 \begin{equation}\label{rel}
\trh = 0.138\sqrt{\frac{\Mcl\rh^3}{G}}\frac{1}{\mmean\psi\ln\Lambda}.
\end{equation}
Here $\mmean$ is the mean mass of the stars and all stellar remnants, which we initially set to $\mmean=\mmeanst\simeq0.638\,\msun$, which is found for a \citet{Kroupa2001} IMF between 0.1 and $100~\msun$. During the evolution $\mmean = \Mcl/\Ncl$, where $\Ncl$ is the initial number of stars which we assume to be constant throughout the evolution, which is accurate in the case of  100\% BH retention and no pair-instability SNe, and slightly overestimates the number of stars once BH ejection occurs and at later times when the turn-off mass drops.  Then, $\ln\Lambda$ is the Coulomb logarithm, which  depends weakly on $\Ncl$, and we therefore use a constant $\ln\Lambda=10$. 

The quantity $\psi$ depends on the mass spectrum within $\rh$, which is usually assumed to be $\psi=1$, applicable to systems of equal mass but in reality $\psi$ can be between $1.5-2$ for GC like mass functions \citep{1971ApJ...164..399S, Kim1998} and $30-100$ for clusters that just formed \citep{2010MNRAS.408L..16G}. In \emacss, a time-dependent $\psi$ was adopted to include the effect of the loss of massive stars by stellar evolution on the relaxation process. However,  \emacss\ was compared to $N$-body models without BHs \citep{Baumgardt2003} and here we want to include the effect of BHs on $\psi$. We therefore search for an expression with a dependence on the properties of the BH population. 

  \citet{1971ApJ...164..399S} derive an expression for $\psi$ for multi-component systems, under the assumption of equipartition (their equation~24): $\psi = \langle \mall^{5/2}\rangle/\langle \mall\rangle^{5/2}$. For our two components model this can be written as: $\psi= (\mst^{3/2}\Mst + \mbh^{3/2}\Mbh)\Ncl^{3/2}/\Mcl^{5/2}$. For $\Mst\simeq\Mcl$ and $\mbh>>\mst$ (such that also the number of BHs is small), this expression for $\psi$ reduces to
\begin{equation}
\psi \simeq 1 +  \frac{\Mbh}{\Mst}\left(\frac{\mbh}{\mst}\right)^{3/2}.
\label{eq:psispitzer}
\end{equation}
The second term on the right-hand side is Spitzer's parameter $S$ \citep{Spitzer1969}, whose value determines whether a system can achieve equipartition ($S\le0.16$). For $\Mbh/\Mst=0.05$ and $\mbh/\mst\simeq20$ \citep[see e.g.][for the case of $\omega$ Cen]{Zocchi2019}, we find $S\simeq4$. This is larger than the criteria for a two-component system to achieve equipartition.  Clusters with such a BH population are therefore not expected to achieve equipartition in the centre \citep[see also][who show this with $N$-body simulations of star clusters with BHs]{Peuten2017}. If we instead of equipartition assume that all members have the same velocity dispersion, then $\psi \simeq 1 + (\Mbh/\Mst)(\mbh/\mst)$, i.e. the index $3/2$ in equation~(\ref{eq:psispitzer}) reduces to 1. 
 An additional complexity is that the values of $\Mbh$ and $\Mst$ in equation~(\ref{eq:psispitzer}) apply to the conditions within $\rh$, where BHs are contributing more to the mass than for the cluster as a whole.

We therefore adopt a  functional form inspired by equation~(\ref{eq:psispitzer}) and introduce a free parameter
\begin{equation}
\psi=1+a_1{\fbh\over 0.01}, 
\label{eq:psi}
\end{equation}
where $\fbh=\Mbh/\Mcl$ is the fraction of the total cluster mass that is in BHs.
With this expression, we have included the dynamical feedback from the BHs on the relaxation process of the cluster. 
 In Section~\ref{ssec:nbody} we {  show that the  evolution of a cluster can be well described by a constant $a_1$ of order unity and } that it is important to include this dependence of $\psi$ on $\fbh$ , as for $\fbh\simeq0.05$ and $a_1\simeq1$, we find $\psi\simeq6$, which has a significant effect on the relaxation process. 
Clearly, relaxation is more important in the early stages when 
$\psi$ is large, while $\psi$ decreases in time because of the ejection of BHs,  narrowing the mass spectrum.  We note that $a_1$ could then also become smaller, because $\mbh/\mst$ decreases as the most massive BHs are ejected first, but in this first version of the model we continue with the simple linear dependence of $\psi$ on $\fbh$ (i.e. equation~\ref{eq:psi}) and reserve a dependence on the BH mass function for future improvements.

In clusters with BHs, the heating in the balanced evolution phase is done by the BHs \citep{Breen2011}, in the form of a BH binary that heats the surrounding BHs, which in turn efficiently transfer the heat to the stars because of the large mass ratio $\mbh/\mst$. 
The dynamical  hardening of BH binaries is an energy source which complies with H\'{e}non's principle \citep{1975IAUS...69..133H}. In this,  the rate of energy generation 
in the core  is set  by the maximum 
heat flow that can be conduced from the core to the
rest of the cluster through two-body
relaxation, and we can therefore relate the heat generation in the core to the  cluster global  properties \citep{1961AnAp...24..369H,2011MNRAS.413.2509G,2013MNRAS.432.2779B}
\begin{equation}
\dot{E} = \zeta\frac{|E|}{\trh}, 
\label{eq:Edot}
\end{equation}
where $E\simeq-0.2G\Mcl^2/\rh$ is the total energy of the cluster, with  the constant $\zeta\simeq0.1$ \citep{1961AnAp...24..369H,2011MNRAS.413.2509G,Alexander2012}. 
{  This definition of the energy excludes the (negative) energy stored in binaries and this energy is therefore sometimes  referred to as the `external energy' \citep{HeggieAarseth1992,GierszHeggie1997}. If heating is done by binaries and in the absence of other changes to $E$ (e.g. stellar evolution, Galactic tides, etc.), negative energy flows into binaries in the centre at a rate $-\dot{E}$, while $E$ increases at a rate $+\dot{E}$ (equation~(\ref{eq:Edot}). As a result, the cluster expands and loses  mass via BH escapers from the core, see e.g. \citet{1984ApJ...280..298G}.}
Note that the efficiency of heat conduction within the cluster is included in $\trh$ via $\psi$, {  in the sense that $\dot{E}$ is higher in clusters with high $\fbh$.}

\citet[][]{2013MNRAS.432.2779B} showed that BHs power the relaxation process, and using theory and  $N$-body simulations they showed that this implies that the mass evolution of the BH population can
be coupled to the properties of the cluster. The explanation for this is, in short, as follows: a BH binary hardens until it ejects itself as the result of the recoil in the last encounter. If all BHs have the same mass, the BH binary ejects on average $\sim4$ BHs before ejecting itself  \citep[see][]{1984ApJ...280..298G}. This allows us to couple the mass-loss rate of BHs to the energy generation rate, which itself is coupled to the total $E$ and $\trh$ of the cluster (equation~\ref{eq:Edot}), such that \citep[][]{2013MNRAS.432.2779B}
\begin{equation}\label{eq:mbhej}
\Mbhdot=
\begin{cases}
0,&t<\tcc{\rm ~or~}\Mbh=0,\\
\displaystyle -\beta  {\Mcl\over \trh}\ , & t\ge\tcc{\rm~and~}\Mbh>0.
\end{cases}
\end{equation}

From the definition of $E$ {  (given just under equatioin~\ref{eq:Edot})} and the assumption of virial equilibrium we find that $\dot{E}/|E| = -2\Mcldot/\Mcl + \rhdot/\rh$, such that the expansion rate as the result of relaxation is
\begin{equation}
\rhdotrlx=\zeta  {\rh\over \trh} + 2\frac{\Mcldot}{\Mcl}\rh\ .
\end{equation}
{  Both stellar mass loss and BH ejection contribute to the mass loss of the cluster, such that} 

\begin{equation}
\Mcldot=\Mstdotsev+\Mbhdot \ .
\end{equation}
{  Quantifying the  interplay between these two mass-loss terms on the energy evolution is beyond the scope of this work, and we simply quantify the rate of BH ejection by fitting $\beta$ to the $N$-body models. 

The final expression for the half-mass radius evolution is then}
 \begin{equation}
\rhdot=
\begin{cases}
\rhdotsev, & t<\tcc,\\
 \rhdotsev + \rhdotrlx , & t\ge\tcc.
\end{cases}
\end{equation}
With all the expressions for the derivatives, we can now numerically solve a set of coupled ordinary differential equations to obtain solutions for $\Mst(t), \Mbh(t)$ (and therefore $\Mcl(t)$) and $\rh(t)$.

We fix the parameters $\tsev=2\,\myr$, $\zeta=0.1$ and determine $\ntrh$, $\beta$, $\nu$ and $a_1$  by fitting the model of this section  to results of direct $N$-body simulations. 
 
 \subsection{Comparison to $N$-body simulations}
\label{ssec:nbody}
To validate the simple model, we  simulate the evolution of clusters with BHs with  direct $N$-body simulations, with different initial cluster densities and BH natal kicks. 
For the calculations we use \nbsix\ (version downloaded on 25 May 2016), which is a fourth-order Hermite integrator with an \citet{AhmadCohen1973}
neighbour scheme \citep{Makino1992, Aarseth1992, Aarseth2003}, 
and force calculations that are accelerated by Graphics Processing
Units \citep[GPUs,][]{NitadoriAarseth2012}. \nbsix\ contains metallicity dependent prescriptions for the evolution of individual stars and binary stars \citep{2000MNRAS.315..543H, Hurley2002}. 

We model 4 different initial densities within $\rh$: $\rhoh=[10^1, 10^2, 10^3, 10^4]\,\msun/\pc^3$, with the initial positions and velocities drawn from a  \citet{Plummer1911} model. For each density we consider two assumptions for the BH kicks: (1) no kicks and  (2) kicks with the same momentum as neutron stars. In the latter, $\vkick$ is drawn from a Maxwell-Boltzmann distribution with  $\sigmak=\sigmans\times1.4\,\msun/\mbh$, and  $\sigmans=190\,\kms$\footnote{{  This is the default value in \nbsix, which is lower than the value of $\sigmans=265\,\kms$ we adopt in the rest of this paper.}}. For $\vesc$ we use equation~(\ref{vesc}), with $f_c=0.68$, {  applicable to the average escape velocity of stars in} a Plummer model.
We use a metallicity of {  $Z=0.0006$} (i.e. $\feh\simeq-1.5$), typical for metal-poor GCs in the Milky Way and all clusters have $N=10^5$ stars initially, drawn from a \citet{Kroupa2001} IMF between 0.1 and 100~$\msun$, such that initially $\mmeanst=0.638\,\msun$. For these parameters the maximum $\mbh\simeq28~\msun$. This is lower than what is found from more recent IFMRs that have been implemented in \nbsix\ \citep{Banerjee2019}, but for the purpose of our model testing this is no concern. In future versions of {  our model} \model\footnote{{  A {\sc python} implementation of the model is available from \href{https://github.com/mgieles/clusterbh}{https://github.com/mgieles/clusterbh}.}} we will compare to a range of \feh\ and IFMRs. All models are evolved until 11.5 Gyr. 

To determine the posteriors of the \model\ parameters, we determine the values of $\Mcl$, $\Mbh$ and $\rh$  from the $N$-body models at 200 equally spaced time intervals between 10 Myr and 11.5 Gyr.  We then define  a log-likelihood

\begin{equation}
\ln\mathcal{L} = -\sum_{k=1}^8\sum_{j=1}^3\sum_{i=1}^{200}  \frac{(D_{ijk}-M_{ijk})^2}{(\delta D_{ijk})^2},
\end{equation}
where $D_{ijk}$ is a $8\times3\times200$ array with the $N$-body data of the  simulations, and $M_{ijk}$ is a similar sized array with the \model\ results. {  The index $k$ loops over the 8 $N$-body models, the index $j$ over the 3 physical quantities ($\Mst, \rh, \Mbh$) and the index $i$ over 200 equally spaced time steps in the range 10 Myr - 11.5 Gyr.} We assume each $N$-body data value has an associated uncertainty $\delta D_{ijk} = 0.05 D_{ijk}$, which captures the fact that cluster parameters evolve with some noise in time and cluster-to-cluster variation. 

The parameters that maximise this log-likelihood are found with  \emcee\ \citep{2013PASP..125..306F}, which is a
pure-\python\ implementation of the Goodman \& Weare's affine
invariant Monte Carlo Markov Chain (MCMC) ensemble sampler
\citep{2010CAMCS...5...65G}. We use 100 walkers and after a few
hundred steps the fit converged. We continued for 1000 steps and in
the analyses we use the final walker positions to generate posterior
distributions. The
\python\ implementation of \emcee\ makes it straightforward to couple
it to \model.

\begin{figure}
\centering
\includegraphics[width=0.5\textwidth]{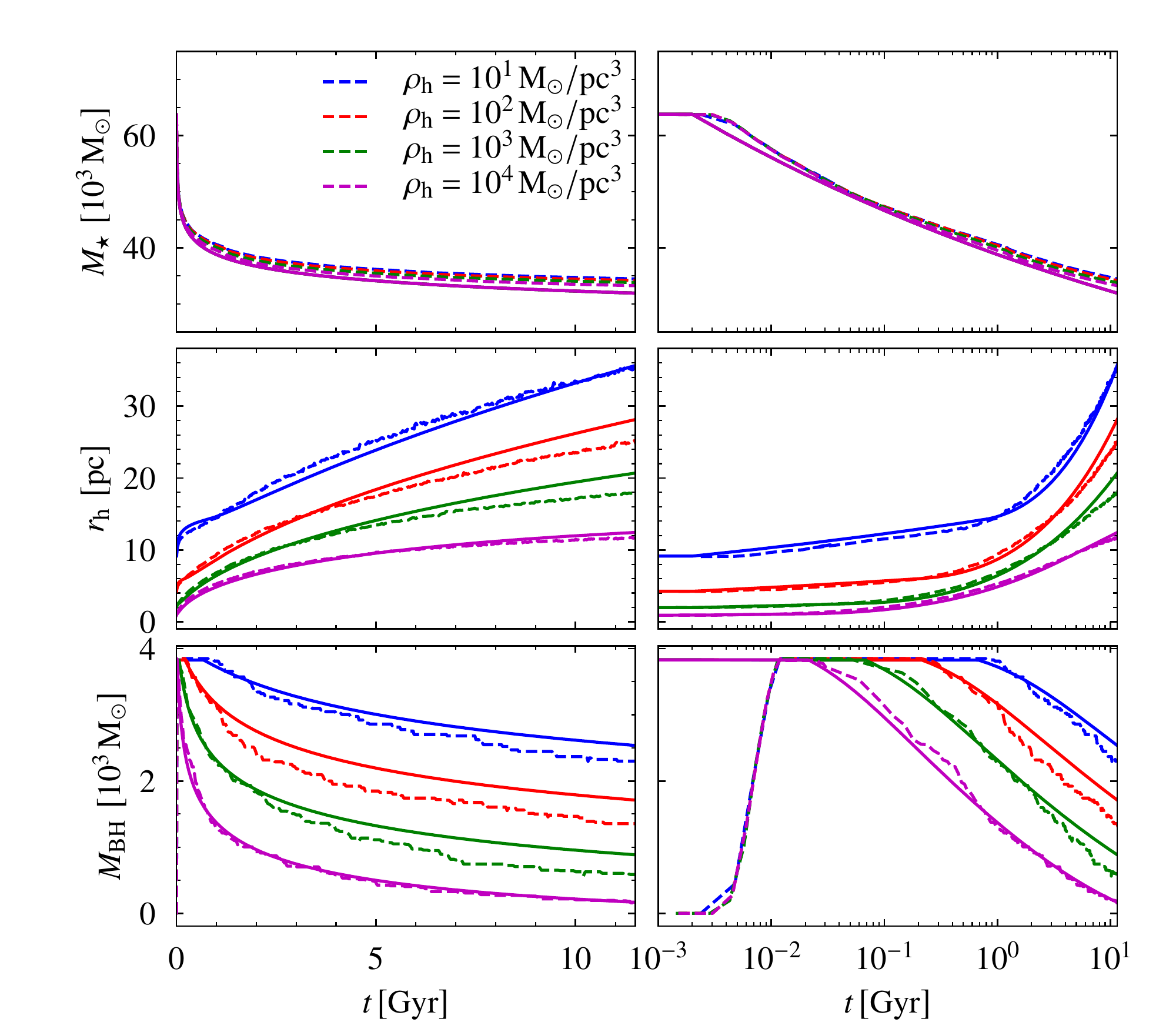}
\caption{$N$-body results of $\Mst$ (top), $\rh$ (middle) and $\Mbh$ (bottom) of isolated star clusters with 100\% BH retention  for different initial densities (dashed lines).  Best-fit \model\ models are shown as full-lines.}
\label{fig:nbody_kick0}
\end{figure}

\begin{figure}
\centering
\includegraphics[width=0.5\textwidth]{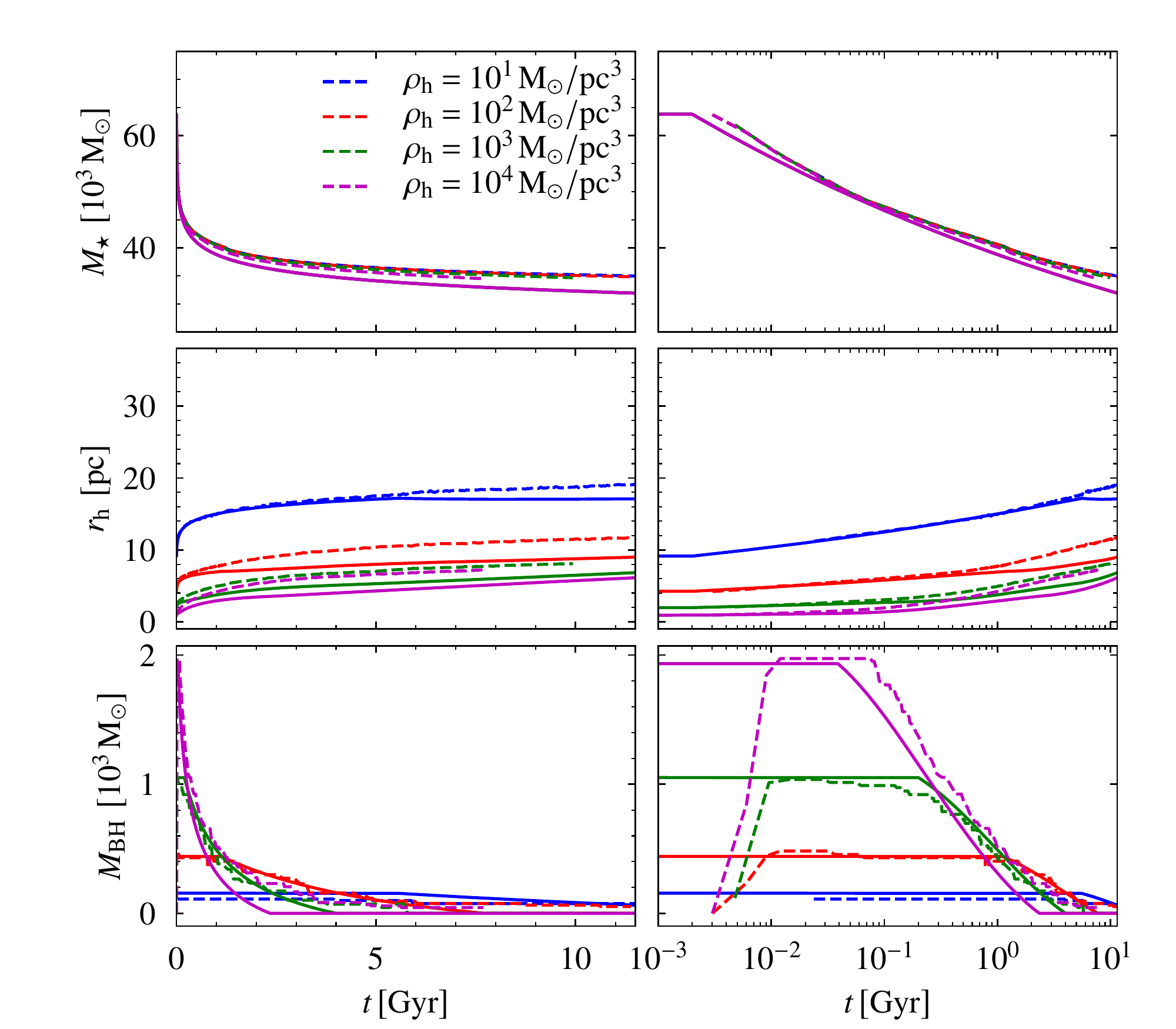}
\caption{$N$-body results of $\Mst$ (top), $\rh$ (middle) and $\Mbh$ (bottom) of isolated star clusters with different initial densities with BH kicks applied (dashed lines).  Best-fit \model\ models are shown as full-lines.}
\label{fig:nbody_kick1}
\end{figure}

For the parameters we find {  $\ntrh = 3.21$, $\beta = 2.80\times10^{-3}$, $\nu = 8.23\times10^{-2}$ and $a_1 = 1.47$}. 
Because the uncertainties on the $N$-body data are artificial, we do not report the uncertainties in the parameters. Instead, we compute for each $N$-body data point the fractional difference with the best-fit model and sort all values. We find that {  68\% of the $N$-body data points are reproduced within 12.8\%}.
Given the simplicity of our model, we consider this accuracy satisfactory. The \model\ results are shown in Fig.~\ref{fig:nbody_kick0} and Fig.~\ref{fig:nbody_kick1} for 100\% BH retention and for momentum conserving BH kicks, respectively. We note that when fitting the model to individual models, better agreement can be obtained, but we are here aiming to obtain model parameters that describe the evolution of all 8 $N$-body models. We note that \model\ slightly over-predicts rate of BH escape for low $\Mbh$ {  (see bottom panels of Fig.~\ref{fig:nbody_kick1})}, which is likely the result of that fact that we did not include a dependence on $m/\mst$ in $\psi$ (equation~\ref{eq:psi}).  Including this would reduce $\vert\Mbhdot\vert$ because when the BH population is about the disappear, the average BH masses are lower than in the early evolution. The good resemblance between the overall evolution of all parameters in the 8 models between the detailed $N$-body models and \model\ confirms that the model does a good job.

\section{Black hole binary evolution}\label{analy}
According to H\'{e}non's  principle 
the flow of energy through the half-mass radius is independent of the precise mechanisms for energy production within the core.
We assume in what 
follows that the heat 
is supplied by the hard  binaries in the core of the BH subsystem  \citep[e.g.,][]{2013MNRAS.432.2779B}.
Assuming that most of the heating is produced by one binary in the core of the star cluster,
we can then relate the binary hardening rate 
to the rate of energy generation
\begin{align}
\dot{E}_{\rm bin} ={\dot{E}} \ .
\label{eq:edotbin}
\end{align}
where $E_{\rm bin}=Gm_1m_2/2a$,
with $a$ the binary semi-major axis
and $m_1$ and $m_2$ the mass of the 
BH components.
In this section we use equation\ (\ref{eq:edotbin})  to  derive the merger rate and eccentricity distributions of merging BH binaries 
produced through binary-single interactions.  To achieve this, we
first need to specify the dynamical mechanisms
that lead to the mergers.

The merger of a BH binary through (strong) binary-single encounters in a dense star cluster following its formation and dynamical hardening
can occur  in three different ways \citep[e.g.,][]{2006ApJ...640..156G,2018PhRvL.120o1101R,2018PhRvD..97j3014S}:  (i) the merger occurs in between 
binary-single encounters while the binary is still bound to its parent cluster (hereafter, we will refer to these mergers
as \emph{in-cluster inspirals});
 (ii) a  merger occurs during a binary-single (resonant) encounter  as two
 BHs are driven to a short separation such that GW radiation will lead to their merger
before  the next intermediate binary-single state is formed
  (hereafter,  \emph{ GW {\it captures}}); and (iii) 
the binary merges after it has been ejected from its parent cluster.
In what follows we 
derive  an analytical model, \modelc 
\footnote{A \fortran\ 
 implementation of these libraries is available from \href{https://github.com/antoninifabio/BHBdynamics}{https://github.com/antoninifabio/BHBdynamics}.}, 
to compute the rate and eccentricity distributions of merging BH binaries  
that are produced through the mechanisms (i), (ii) and (iii).
This model can then be easily coupled to \model\ in order to 
compute such distributions for and evolving model
of a star cluster (Section\ \ref{AI}).
Hereafter, the combination of our two models is referred  to as   \modelb .

\subsection{Merger fractions}
\subsubsection{In-cluster inspirals.}\label{3rec}
The binary-single interactions  in the cluster core
lead to a decrease in the binary semi-major axis until 
the binary evolution becomes dominated by GW energy loss.
The timescale, $t_3$, over which a binary will 
encounter  another 
BH in the cluster core can be obtained by noting
that { binaries release on average 20\% of their binding energy to the passing BH in an encounter, such that}
$\dot{E}_{\rm bin}\simeq 0.2{E}_{\rm bin}/t_{3}$ \citep{Heggie2003}, which leads to
the relation
\begin{align}
t_{3}\simeq 0.2 {Gm_1m_2\over2a}\dot{E}_{\rm bin}^{-1}\ .
\label{eq:thard}
\end{align}
At later times, the evolution of the binary semi-major axis and eccentricity  is 
described by the orbit averaged evolution equations \citep{Peters1964}:
\begin{align}\label{peta}
\dot{a}_{\rm GW}=
-{64\over 5}\frac{G^3m_1m_2m_{12}}{c^5a^3\ell^{7}} \left(1+{73\over24}e^2+\frac{37}{96}e^4\right)\ 
\end{align}
\begin{align}\label{edot}
\dot{e}_{\rm GW}=-{304\over 15}{G^3m_1m_2m_{12}\over c^5a^4\ell^{5}}
\left(e+{121\over304}e^3\right)\ ,
\end{align}
with $\ell\equiv (1-e^2)^{1/2}$ {  the dimensionless angular momentum}, $e$  the binary eccentricity,
and $m_{12}=m_1+m_2$.
We define the  corresponding merger timescale 
as $t_{\rm GW}\equiv a/\left|\dot{a}_{\rm GW}\right|$.

The transition from the dynamical
to the GW regimes  occurs at 
$t_{\rm GW}\lesssim t_{3}$, or equivalently when for a given $a$ the eccentricity is larger than
 \begin{align}\label{gw}
\ell<\ell_{\rm GW}\simeq
1.3\left[
{G^4\left(m_1m_2\right)^2m_{12}
\over
c^5\dot{E}_{\rm bin}}
\right]^{1/7}a^{-5/7} \ ,
\end{align} 
{  where $\dot{E}_{\rm bin}$ is related to the  properties of the cluster through equation\ (\ref{eq:Edot}),}
and we
have taken the relevant limit  $e\rightarrow 1$.
For $\ell\lesssim \ell_{\rm GW}$ the evolution  becomes dominated by
GW energy loss and the binary inspirals approximately as an isolated system.

During  a single interaction with a cluster member of mass $m_3$
the semi-major axis of the binary decreases from $a$
to $\epsilon a$. Then energy and momentum conservation imply
that the binary  will experience a  recoil kick 
 $v_{\rm bin}^2 \simeq G\mu {m_3\over
  {m_{123}}} \left[1/(\epsilon a)-1/a\right] = \left(1/\epsilon-1\right) G{m_1m_2\over m_{123}}  {q_3/a}$, where $\mu=m_1m_2/m_{12}$, $m_{123}=m_{12}+m_3$,
$q_3=m_3/m_{12}$,  
and we have assumed that in the interaction
the binding energy of the binary increases by the fixed fraction
$\left(1/\epsilon-1\right)=0.2$, which implies $\epsilon=0.83$ \citep[e.g.,][]{1987degc.book.....S,
2010ARA&A..48..431P}.
Taking $v_{\rm bin}= v_{\rm esc}$, with $v_{\rm esc}$ the escape velocity from the cluster,
 we  obtain the limiting semi-major axis below which a three
body interaction will eject the binary from the cluster \citep{2016ApJ...831..187A}:
\begin{align}\label{aej}
a_{\rm ej}= \left({1\over \epsilon}-1\right)G {m_1m_2\over m_{123}} {{q_3}}/{v_{\rm esc}^{2}} .
\end{align}

Given a criterion for when a binary enters the GW regime (equation~\ref{gw}) and for when
a binary is ejected from the cluster (equation~\ref{aej}), we can now compute
how probable it is for a binary  to attain $\ell<\ell_{\rm GW}$ as it hardens through 
binary-single interactions in the cluster core.

{  We assume that during each binary-single encounter, the binary 
receives a large angular momentum kick such  that the phase
space is stochastically scanned and approximately  uniformly covered by the periapsis values \citep[e.g.,][]{2012arXiv1211.4584K}.
Thus,  during a typical encounter the BH binary semi-major axis will shrink while the orbital eccentricity will be randomly drawn from the thermal distribution
$p(>e)=1-e^2= \ell^2$.
It follows that the
 probability that a binary merges
between two successive  binary-single encounters  is
\begin{align}\label{prob0}
p_{\rm GW}(a)= \ell^2_{\rm GW}\ .
\end{align}
}

 The total probability that a binary merges in between
its binary-single interactions {  ($P_{\rm GW}$)}  is then
obtained by integrating the differential merger probability 
per binary-single encounter, $\dr{P}_{\rm GW}= p_{\rm GW}\dr N_{3}$, over the
total number  of binary-single interactions experienced by the binary.
Noting that  $\dr a/\dr N_{\rm 3}=(\epsilon-1)a$, this leads to \footnote{A similar expression can be found
 in \cite{2018PhRvD..97j3014S}.}
\begin{align}\label{Prob}
{P}_{\rm GW}(a_{\rm m}) =
\int_{a_{\rm h}}^{a_{\rm m}} 
{1\over \epsilon-1}
{\ell_{\rm GW}^2 }({a}) {\dr {a}\over {a}} \simeq
{7\over 10}{1\over 1-\epsilon}\ell_{\rm GW}^2(a_{\rm m}),
 \end{align}
 where $a_{\rm h}$ is the semi-major axis of a binary at the hard/soft boundary {  (i.e. when it forms)},
 $a_{\rm m}$ is  the  minimum semi-major axis value that can be attained 
 by any binary during the hardening sequence, and
in deriving the last expression we have assumed $a_{\rm h}\gg a_{\rm m}$\footnote{We note that
the total merger probability before a state $n$
of the hardening sequence is reached is 
$P(n)={\prod^{ n}_{i=0}}{p}_{i}$, where ${p_{i}}$ is the probability that the binary  
 merged at state $i$. This gives 
$P(n)\approx {7\over 10}{1\over 1-\epsilon}p_n$ only
at leading order
\citep{2019arXiv190102889S}.}.
The value of $a_{\rm m}$ is
  \begin{align}
a_{\rm m}\simeq   \max \left(a_{\rm ej};\ a_{\rm GW}\right)\ ,
\end{align}
with $a_{\rm GW}$
  the semi-major axis at which  
  the merger probability in between binary-single interactions is one,  i.e., ${P}_{\rm GW}(a_{\rm GW})=1$,
  such that
  \begin{align}\label{agw}
a_{\rm GW}\simeq1.1\left( { 1-\epsilon}\right)^{-7/10} \left[
{G^4\left(m_1m_2\right)^2m_{12}
\over
c^5\dot{E}_{\rm bin}}
\right]^{1/5} .
  \end{align}
The binary-single interactions will terminate either because the binary is
ejected from the cluster or because the binary  merges in between binary-single interactions.

From equations~(\ref{Prob}) and (\ref{gw}) it can be seen that {  the probability that a binary will merge inside the cluster before ejection is ${P}_{\rm GW}\propto v_{\rm esc}^{20/7}$ for $a_{\rm m}=a_{\rm ej}$}, where we neglect the weak dependence of $\ell_{\rm GW}$ on $\dot{E}_{\rm bin}$. Note that  for $a_{\rm m}=a_{\rm GW}$ 
all  merges occur inside the cluster and ${P}_{\rm GW}=1$.
The probability that a binary  will merge  inside its parent cluster before ejection is possible is
therefore related   to the host  physical properties through  
its escape velocity
\begin{align}\label{vesc}
\vesc\simeq\ 50\,{\rm km\ s^{-1}}  M_5^{1/3} \rho_5^{1/6}{f_{c}},
\end{align}
where we expressed the result in terms of $M_5 = M_{\rm cl}/10^5\,\msun$ and $\rho_5 = \rhoh/10^5\,\msun/\pc^3$, with $\rhoh = 3M_{\rm cl}/(8\pi\rh^3)$  
the average density within $\rh$.
The coefficient $f_c$ in the latter equation takes into account the dependence
of the escape velocity on {  the location within the cluster and} the concentration of the
cluster, i.e., $c=\log(r_{\rm t}/r_0)$ with $r_{\rm t}$  the cluster
truncation radius and $r_0$ the King (or, core) radius. Below we simply set $f_{c}=1$ which
corresponds to {  the centre of} a King model with concentration parameter $W_0\simeq7$.
The other factors affecting whether a binary can merge within the cluster are the
masses of the participants in the interactions and therefore the mass function of the black holes
and stars in the cluster. 
A simple model for the evolution of the BH mass function  is
introduced later in Section\ \ref{bhmass}, where we make the assumption
that $m_1=m_2=m_3$ and all are equal to the most massive BH in the cluster at that time.

\subsubsection{Gravitational wave captures}
In the previous section we did not consider 
 mergers that occur by `direct capture' during a three-body resonant encounter.
Although such mergers are expected to be only a small fraction  of the
total \citep[$\lesssim 10\%$;][]{2006ApJ...640..156G,2018PhRvD..97j3014S}, they are  interesting because they might have  residual eccentricities in the LIGO-Virgo frequency band.
For this reason we include them in the analysis below.

Following \cite{2018PhRvD..97j3014S}, we divide each resonant encounter into  a number $N_{\rm IS}$ of intermediate binary-single states,
where an intermediate BH binary is formed with a bound companion.
Using a large set of three-body scattering experiments \citet{2018PhRvD..97j3014S} finds $N_{\rm IS}\simeq 20$,
which is the value we adopt throughout this paper.
Moreover, we define the characteristic angular momentum  below which two of the BHs can undergo a GW merger during an
intermediate binary-single state, $\ell_{\rm cap}$, as that for which the GW energy loss integrated 
over one periapsis passage becomes of the order the orbital energy of the binary. One finds that at \citep{Samsing2014,2018PhRvD..97j3014S}
 \begin{equation}\label{lcapt}
\ell<\ell_{\rm cap}\simeq h \left(  R_{\rm S}\over a \right)^{5/14}\ ,
\end{equation}
a GW merger will occur before the next  intermediate binary-single state is formed.
In the previous expression, $R_{\rm S}=2Gm_{12}/c^2$, and
the constant $h$ is of order unity.

Assuming that the eccentricity distribution of the intermediate 
state binaries follows a thermal distribution, 
the probability per  encounter that a GW capture will occur is  
\begin{equation}\label{prob01}
p_{\rm cap}(a)= N_{\rm IS} \ell^2_{\rm cap}\ .
\end{equation}

By using that the differential merger probability per encounter is $\dr{P}_{\rm cap}= p_{\rm cap}\dr N_{3}$,
and integrating over all binary-single encounters one finds the total merger probability via GW capture:
\begin{align}
{P}_{\rm cap}(a_{\rm m})= N_{\rm IS}\int^{a_{\rm m}}_{a_{\rm h}}{1\over {\epsilon -1}}
\ell^2_{\rm cap}({a}) {\dr {a}\over {a}}\simeq
N_{\rm IS}{7\over 5}{1\over {1-\epsilon }} \ell^2_{\rm cap}(a_{\rm m}),
\end{align}
where as before we have used that $a_{\rm h}>>a_{\rm m}$.
If GW {captures} are included, then $a_{\rm m}$ has to be redefined 
as the semi-major axis where the total in-cluster merger probability, 
${P}_{\rm in}={P}_{\rm GW}+{P}_{\rm cap} $, is unity.
From this latter condition we find:
\begin{align}\label{agw2}
a_{\rm GW}= \left({\sqrt{N_{\rm IS}^2g_{\rm cap}^2 +{10\over 7}(1-\epsilon)g_ {\rm GW}}
-N_{\rm IS}g_{\rm cap}\over g_{\rm GW}}\right)^{-7/5}
\end{align}
where $g_{\rm cap}=h^2R_{\rm S}^{5/7}$
and $g_ {\rm GW}=1.7\left[ {G^4\left(m_1m_2\right)^2
m_{12}
\over c^5\dot{E}_{\rm bin}} \right]^{2/7}$. 
A value for $h$ was derived by fitting  the GW capture fractions in the three-body scattering experiments of Table 1 in \citet{2006ApJ...640..156G}.
We find the best fit value to be $h=1.8$, 
which we adopt in what follows.
Note that  in the limit $R_{\rm S}\rightarrow 0$,  equation\ (\ref{agw2})
becomes equation\ (\ref{agw}).

\subsubsection{Ejected binaries}
A BH binary ejected at a (look-back) time $\tau$ from its parent cluster will merge within the present time
if its angular momentum satisfies the relation 
\begin{align}
\ell < \ell_{\rm H}&\simeq 1.8\left[\frac{G^3m_1m_2m_{12}}{c^5}  \tau
\right]^{1/7}a^{-4/7}\\
&\simeq
\left(0.07{\rm AU}\over a\right)^{4/7} \left({m_1m_2m_{12}\over
10^3\,\msun^3} {\tau\over10\,\rm Gyr}
\right)^{1/7} \nonumber \ ,
\end{align}
which was derived by setting $t_{\rm GW}<\tau$ in the limit of large eccentricities.

Detailed Monte Carlo  simulations of the secular evolution of massive star clusters show
that dynamically ejected BH binaries have eccentricities  which 
are well described by a thermal distribution \citep[e.g.,][]{2016ApJ...830L..18B}. 
Thus, the  probability 
that an ejected binary will merge on a timescale shorter  than $\tau$
 is  
\begin{equation}
p_{\rm ex}(a_{\rm ej})=\ell_{\rm H}^2(a_{\rm ej}) \ .
\end{equation}
Finally, the total probability that a BH binary  will merge
outside its parent cluster is obtained by multiplying the
probability that the binary did not merge inside the cluster (i.e., $1-{P}_{\rm in}$),
by the probability that the binary will merge after being ejected:
\begin{equation}\label{pex}
{P}_{\rm ex}(a_{\rm m},a_{\rm ej})=\left[1-{P}_{\rm in}(a_{\rm m})\right]p_{\rm ex}(a_{\rm ej}).
\end{equation}
{  Note that given the definition of $a_{\rm m}$ above, we have that $0\le {P}_{\rm in}\le 1$. From equation\ (\ref{pex})
it then follows that $0\le {P}_{\rm ex}\le 1$.}

\subsection{Merger rates}
{  In balanced evolution,  the binary  formation rate is  
equal to the binary ejection rate and can therefore be expressed as
 the  BH mass ejection rate of equation~(\ref{eq:mbhej}) divided by
the total mass ejected by each binary
\begin{align}\label{rateF}
\Gamma_{\rm bin}&\simeq - \frac{\dot{M}_{\rm BH}}{ m_{\rm ej}},\nonumber\\
&\simeq
\frac{0.4}{\rm Myr}\frac{60\,\msun}{ m_{\rm ej}}{\beta\over 2.8\times 10^{-3}} \frac{\psi}{5}\frac{\mmean}{0.4\,\msun}\frac{\ln\Lambda}{10}
 \rho_5^{1/2} \ .
\end{align}}
Note that because the heating produced in the core is only determined by
the global properties of the cluster,  the binary formation rate  equation\ (\ref{rateF}) is  independent of the cluster core properties
 and the number of binaries in the core \citep[see also][]{2019MNRAS.486.5008A}.

Noting that the condition for the recoil velocity of the
interloper to be larger than $v_{\rm esc}$
is $a\leq a_{\rm 3}=a_{\rm ej}/q_3^2$, then the total 
mass a BH binary ejects, $m_{\rm ej}$, is 
equal to the binary mass plus the total mass
ejected through  the binary single interactions experienced  from $a_3$ to $a_{\rm end}$,
where $a_{\rm end}$ is the semi-major axis at which the sequence of binary-single interactions
terminates, such that
\begin{align}\label{rateM}
m_{\rm ej}=m_{12}+\int^{a_{\rm end}}_{a_{3}}{m_3\over{\epsilon-1}} {\dr {a}\over {a}}\simeq
m_{12}+{m_3\over {1-\epsilon}} \ln \left({a_{\rm ej} \over a_{\rm m}} q_3^{-2} \right) \ .
\end{align}
In the last expression we have set $a_{\rm end}=a_{\rm m}$,
which is  only correct for ejected binaries. {  For in-cluster mergers
$a_{\rm end}$ is a distribution of values
because a binary has a finite probability to merge 
at any point between 
$a_{\rm h}$ and $a_{\rm m}$.}
 However,
equation~(\ref{rateM}) is still a reasonable  approximation given that
 $m_{\rm ej}$
depends on ${a}_{\rm end}$ only through the
logarithmic term, and 
 equation\ (\ref{Prob}) shows that for in-cluster mergers the
distribution of ${a}_{\rm end}$  has a median value of 
$1.6a_{\rm m}$. 
We note another difference: for in-cluster mergers  the binary is not 
ejected dynamically from the cluster. However, it will be
ejected almost certainly by the relativistic recoil kick
after a merger occurs \citep{2016ApJ...831..187A}.

{  
In  previous studies, 
the  three-body binary formation rate was computed via
 `microscopic' approaches that directly consider the three-body interactions and which introduce a relation to the cluster core properties:
 $\Gamma_{\rm bin}\simeq \sqrt{2}\pi^2G^5n^2\sigma_{\rm c}^{-9}m_{12}^5$,
with $\sigma_{\rm c}$ the velocity dispersion of the BHs and $n$  their number density
in the core \citep[e.g.,][]{Ivanova2005,Morscher2015}. 
In the present framework instead, $\Gamma_{\rm bin}$
 is derived via a H\'{e}non's principle-based (or a `macroscopic') approach and it is 
 therefore only expressed in terms of $M_{\rm cl}$ and $r_{\rm h}$. 
Clearly,
the advantage of our technique is that it does not require any detailed knowledge of the evolution of the cluster core properties which  requires computationally expensive  simulations.
}

{  The  rate $\Gamma_{\rm bin}$ is the rate at which 
binaries harden from an initial semi-major axis $a_{\rm h}$ to a separation, $a_{\rm m}$, where
the GW inspiral phase starts.
Thus, multiplying  $\Gamma_{\rm bin}$ by the merger probability 
 and
integrating over time we obtain the total number of mergers produced within a given time interval. }
For the time-dependent cluster model of Section~\ref{clev},
 the   number 
of BH  binaries that merge within a redshift $z<z_{\rm d}$  is then
\begin{align}
\mathcal{N}&=\int_{0}^{z_{\rm d}}\Gamma_{\rm bin} 
\left[{{P}}_{\rm GW}+{{P}}_{\rm cap}+{{P}}_{\rm ex} \right]
  {\dr {\tau}\over \dr z}\dr z  \nonumber\\
&+\int^{z_0}_{z_d}\Gamma_{\rm bin}
\left[{{P}}_{\rm ex}({\tau})-{{P}}_{\rm ex}({\tau}-\tau_{\rm d}) \right]\
{\dr {\tau}\over \dr z}\dr z \ ,
\label{eqN}
\end{align}
where $z_0$ is the formation redshift of the cluster, and 
$\tau_{\rm d}$ is the look-back time corresponding to a redshift $z_{\rm d}$.
The second term in the right-hand-side of equation~(\ref{eqN}) is the
number of BH binaries which are ejected at redshift $z>z_{\rm d}$ but merge at redshift $z<z_{\rm d}$ (i.e., within the observable volume).
The look-back time and redshift are related through the equation
\begin{equation}
{\dr\tau\over \dr z}={1\over
H_0(1+z)\sqrt{\Omega_{\rm M}(1+z)^3+\Omega_\Lambda}
} \ ,
\end{equation}
and we assume here a $\Lambda$CDM flat cosmology with the $Planck$ values for the cosmological parameters, $H_0 = 67.8 
\rm km\ s^{-1}Mpc^{-1}$ and $\Omega_{\rm M}= 0.308$ \citep{PlanckCollaboration2015}.

Equation~(\ref{eqN}) can be numerically  integrated 
to compute the rate at which merging BHs  are
produced dynamically in a dense star cluster
and can be used  to rapidly explore the
 relationship between the merger rate and a cluster's global properties.

\subsection{Eccentricity distributions}
The distribution of eccentricities of merging binaries at a given gravitational wave 
frequency, $f$, for the three populations of mergers can be approximated  as described in what follows.

For a binary evolving under the influence of GW radiation,
 Peters equation \citep{Peters1964}  is used to relate the  
 binary eccentricity  to its semi-major axis
\begin{align} \label{ferel}
{a_0\ell_0^2\over a \ell^2}=\left(e_0\over e\right)^{12/19}
\left({1+{121\over304}e_0^2} \over
{1+{121\over304}e^2}
\right)^{870/2299} \ .
\end{align}
Using
that the  peak GW frequency $f$ of a binary with semi-major axis $a$ and
eccentricity $e$ is \citep{Wen2003}
\begin{align}\label{pf}
f={1\over \pi}{\sqrt{Gm_{12}\over  a^3}} {\left(1+e\right)^{1.1954}\over  \ell^3}  \ ,
\end{align}
we can rewrite equation~(\ref{ferel}) as
 \begin{align} \label{ferel2}
{\ell_0^2}&={1\over a_0}{\left(Gm_{12}\over \pi^{2}\right)^{1/3}} \left(1\over f \right)^{2/3}
\left(e_0\over e\right)^{12/19} \nonumber \\ 
&\times \left({1+{121\over304}e_0^2} \over
{1+{121\over304}e^2}
\right)^{870/2299}\left(1+e \right)^{0.7969} \ .
\end{align}
 The previous equation can be 
further simplified by noting that in the limit where 
$e_0\approx 1 $, as justified for most dynamically formed binary BH mergers, we can write
 \begin{align}\label{eapp}
\ell_0^2\simeq {1\over a_0}{\left(Gm_{12}\over \pi^{2}\right)^{1/3}}
 F(e, f),
\end{align}
with
\begin{align}
 F(e, f)&=1.14 \left(1\over f \right)^{2/3}\left(1\over e \right)^{12/19} \nonumber \\
 & \times\left(1 \over
{1+{121\over304}e^2}
\right)^{870/2299}{(1+e)^{0.7969}}\ .
\end{align}
An isolated binary with initial semi-major axis $a_0$ and angular momentum   $\ell_0$
will have an eccentricity $e$ at a   GW frequency  $f$.

For in-cluster inspirals,  the
differential probability that 
the inspiral will start from an 
angular momentum 
larger than a given $\ell_0$ 
is $\dr{\mathcal{P}}_{\rm GW}=\left(\ell^2_{\rm GW}-\ell_0^2\right)\dr N_3$.
Integrating this probability over the 
total number
of binary-single interactions from
formation to
merger gives the eccentricity distribution
of merging BH binaries at a GW frequency $f$:
\begin{align}\label{cum1}
{\mathcal{P}}_{\rm GW} &= {{\int_{a_{\rm h}}^{a_{\rm m}} 
{1\over \epsilon -1} \left( {\ell_{\rm GW}^2({a})}-\ell_0^2({a})\right) {\dr {a}\over {a}} }}
\nonumber \\ 
&\simeq{1\over 1-\epsilon} 
{\left[ {{7\over10}\ell_{\rm GW}^2}(a_{\rm m})-  
{{\left(Gm_{12} \over\pi^{2}\right)^{1/3}} {F(e, f)\over a_{\rm m}}} \right] \ .
}  
 \end{align}
For GW captures, the probability that inspiral will start  from an angular momentum 
larger than $\ell_0$ 
is $\dr{\mathcal{P}}_{\rm cap}=\left(\ell^2_{\rm cap}-\ell_0^2\right)\dr N_3$, and the eccentricity distribution at $f$ is 
  \begin{align}\label{cum2}
{\mathcal{P}}_{\rm cap} &=  N_{\rm IS}{{\int_{a_{\rm h}}^{a_{\rm m}} 
{1\over \epsilon -1}\left( {\ell^2_{\rm cap}({a}) }-\ell^2_0({a})\right) {\dr {a}\over {a}} }}
 \nonumber\\ 
&\simeq
{N_{\rm IS}\over 1-\epsilon}{\left[ {7\over 5}{\ell_{\rm cap}^2}(a_{\rm m})-  
{{\left(Gm_{12} \over\pi^{2}\right)^{1/3}} {F(e, f)\over a_{\rm m}}} \right]
 \ .
}  
 \end{align}
Similarly, the  cumulative eccentricity distribution of binaries merging after being ejected from their host cluster at a given GW frequency 
$f$ is 
 \begin{align}\label{cum3}
{\mathcal{P}}_{\rm ex}=
\left(1-{P_{\rm in}}\right)
\left[{{\ell_{\rm H}^2}(a_{\rm ej})-  
{{\left(Gm_{12} \over\pi^{2}\right)^{1/3}} {F(e, f)\over a_{\rm ej}}}  
} \right]\ ,
 \end{align}
{  where the second term inside the parenthesis comes from equation\ (\ref{eapp}) with $a_0$ replaced by $a_{\rm ej}$.}

Finally, integrating equation~(\ref{cum1})-(\ref{cum3}) with respect to time
gives the eccentricity distribution of the entire population of binary BHs
that merge at a redshift $z<z_{\rm d}$  and at a  GW frequency $f$ with an eccentricity less than $e$:
\begin{align}\label{eqNe}
{\mathcal N}_{<e}&=\int_{0}^{z_{\rm d}}\Gamma_{\rm bin} 
\left[{\mathcal{P}}_{\rm GW}+{\mathcal{P}}_{\rm cap}+{\mathcal{P}}_{\rm ex} \right]
  {\dr {\tau}\over \dr z}\dr z \nonumber \\
&+\int^{z_0}_{z_d}\Gamma_{\rm bin}
\left[{\mathcal{P}}_{\rm ex}({\tau})-{\mathcal{P}}_{\rm ex}({\tau}-\tau_{\rm d}) \right]\
{\dr {\tau}\over \dr z}\dr z \ ,
\end{align}
where, as before, the first term in the right-hand-side of the equation represents the number of mergers from binaries that are formed within a redshift $z_{\rm d}$, and  the second term  is the
number of BH binaries which are ejected at redshift $z>z_{\rm d}$ but merge at redshift $z<z_{\rm d}$. We compute later in Section~\ref{ediss} the eccentricity distribution of BH binary mergers  for a wide range of cluster initial conditions.

\begin{figure}
\centering
\includegraphics[width=2.6in,angle=0.]{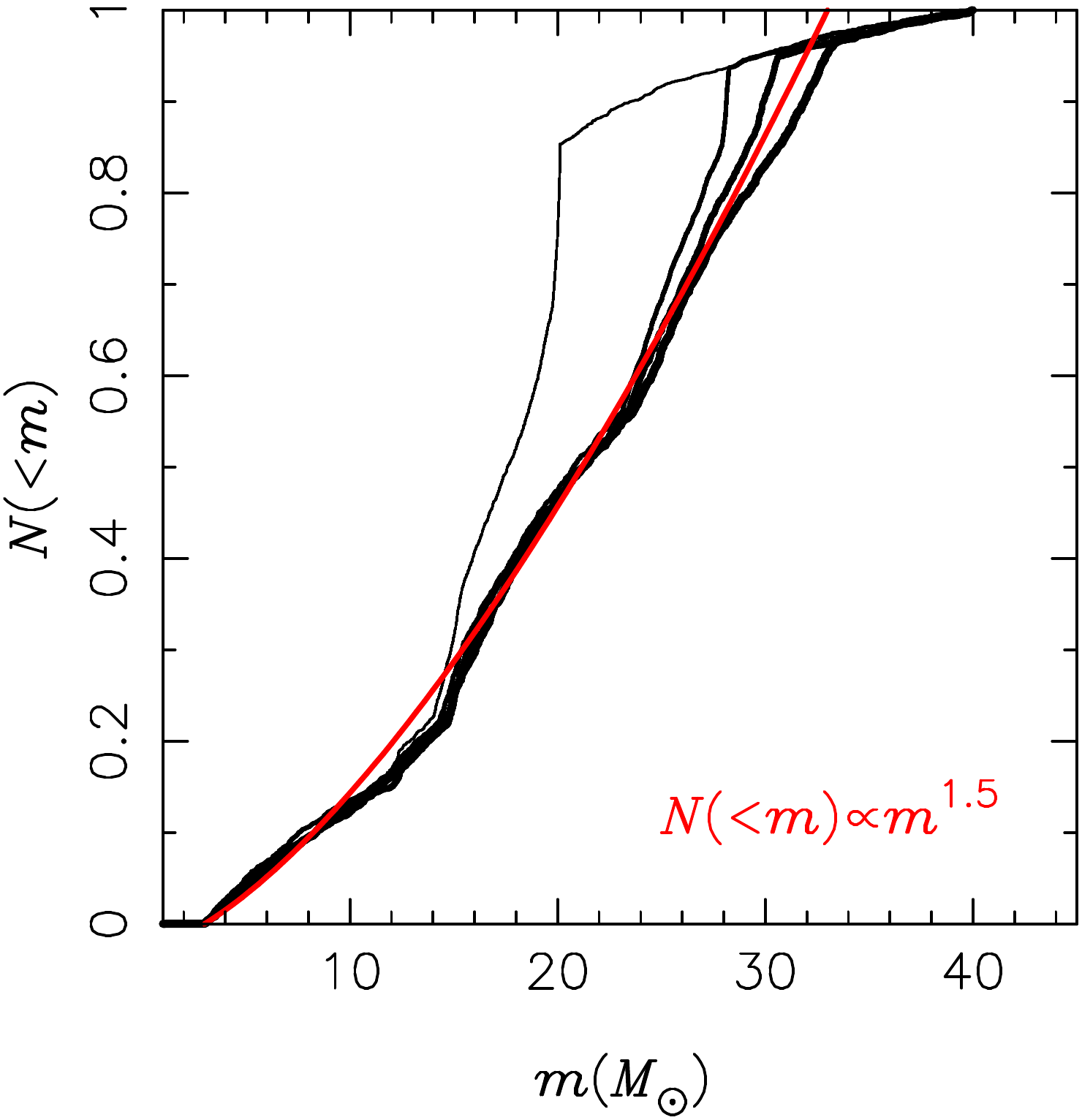}
\caption{Cumulative distribution of the initial BH masses obtained
from the stellar evolution calculations for 
$Z=(0.1, 0.05, 0.025, 0.01)Z_\odot$, where 
line thickness decreases with increasing metallicity.
The red line shows the cumulative distribution 
corresponding to the BH mass function
$\phi\propto m^\alpha$ with $\alpha=0.5$ between 
$m_{\rm lo}=3\,\msun$ and $m_{\rm up}=33\,\msun$
which  is a good approximation
to the numerical results   
for $Z\lesssim 0.05Z_\odot$.
}\label{maf}
\end{figure}

\begin{figure*}
\centering
\includegraphics[width=2.in,angle=270.]{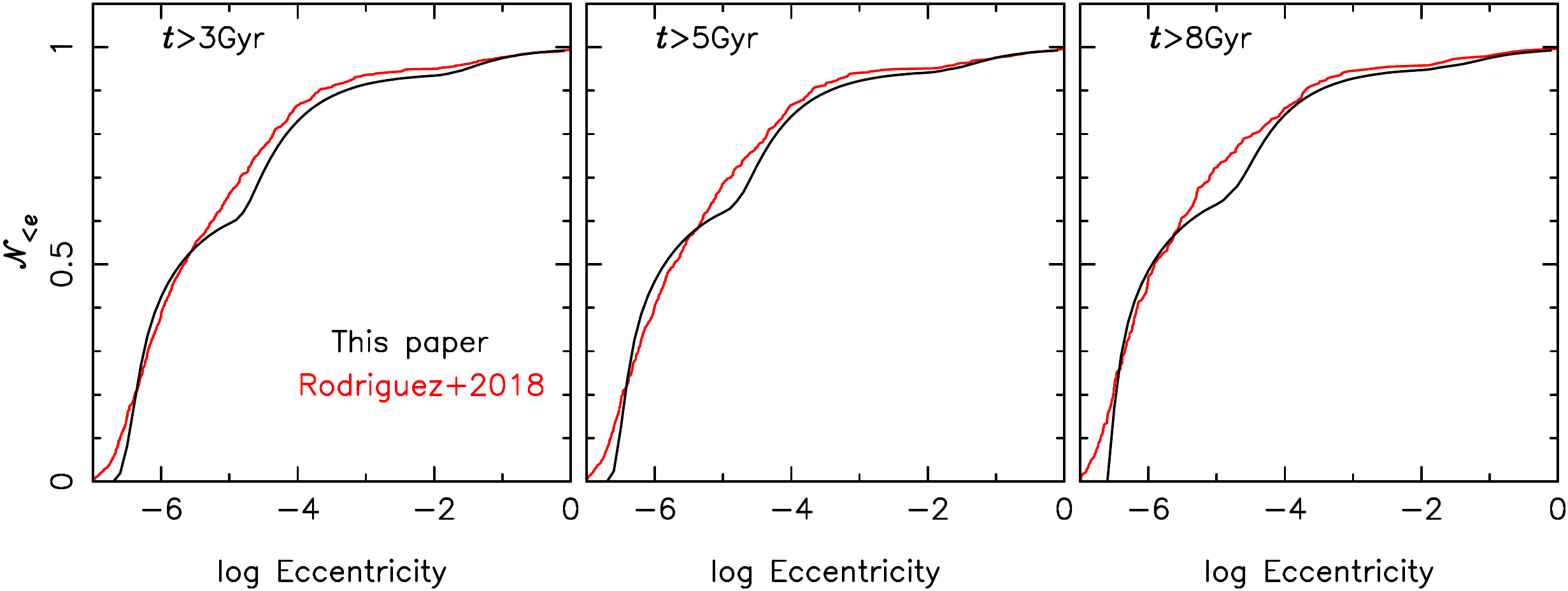} 
\includegraphics[width=2.in,angle=270.]{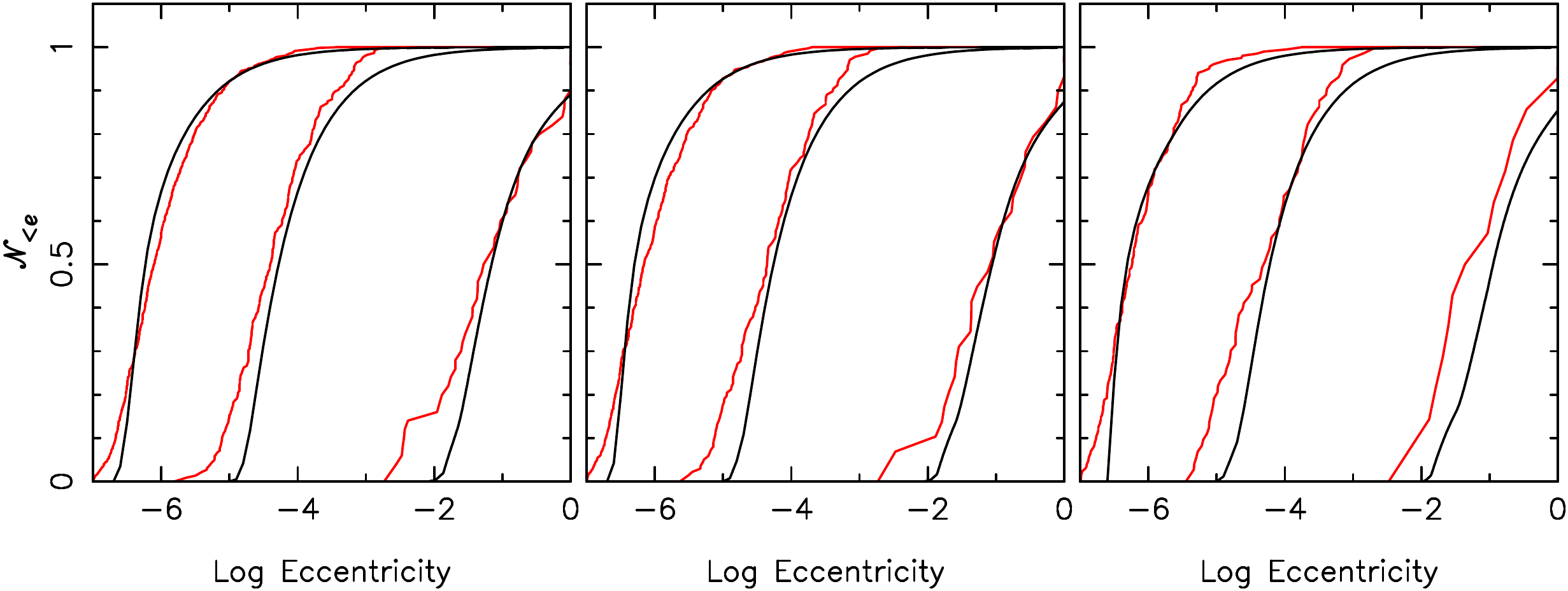} 
\caption{Comparison between the eccentricity distribution of the BH mergers produced in the Monte Carlo simulations of
 \citet{2018PhRvD..98l3005R} and
those obtained with our  method.
Top panels show the cumulative distribution of $e$ for all mergers. In the bottom panels, 
the distributions  of GW captures, in-cluster mergers,  and mergers among ejected binaries are shown separately (from right to left, respectively).
The distributions refer to binaries  that  merge after a given time from the start of the evolution, as
indicated. 
}\label{mccomp}
\end{figure*}

\begin{figure*}
\centering
\includegraphics[width=5.3in,angle=0.]{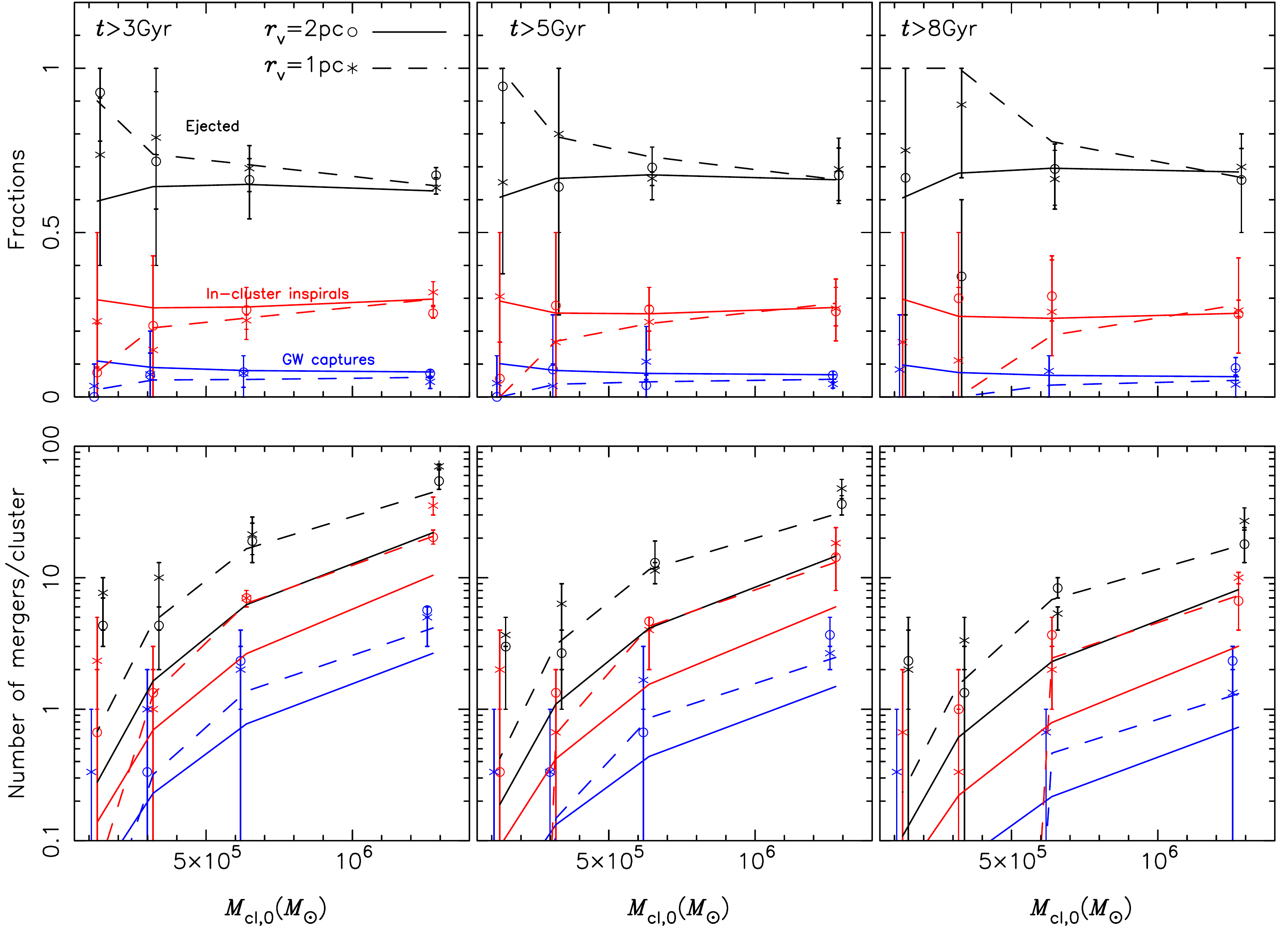} 
\caption{
Fractions and total number
of mergers for each of the three  merger channels separately, at different times
 as a function of initial cluster mass obtained using our method (lines), and
 from the full Monte Carlo models of 
 \citet{2018PhRvD..98l3005R} (symbols). Each point
 represents the average number of mergers from the three 
 metallicities adopted in  the Monte Carlo models.
}\label{mccomp2}
\end{figure*}

\subsection{Black hole masses}\label{bhmass}
For the calculation of the merger rate and eccentricity distributions in \modelc\
we need a model for the time evolution of the BH mass
function. We assume that
  the BHs participating to the
  interactions have the same mass, i.e., 
  $m_1=m_2=m_3$ and use $\mmax$ below to indicate the mass of each of the three BHs. The value of $\mmax$ is by definition restricted to the range of $\phi_0$, i.e. $\mlo\le\mmax\le\mup$.  This  choice is  motivated by 
theory and detailed numerical models
of dense star clusters \citep[e.g.,][]{Sigurdsson1993,Rodriguez2016a}. Because of equipartition of energy, 
 during a
binary-single encounter the two heaviest objects in the interaction are the most likely to be paired together. 
Thus, after a few  interactions the two BH components of a hard binary will  have a similar mass.
Moreover, because of mass segregation, the core densities 
will be dominated by the heaviest objects, so the BHs that the binary will encounter more frequently  will
have a mass comparable to the mass of its components, which are the most massive BHs still present in the cluster.

The cluster  will process  its BH population such
that the mass of the  ejected BHs 
progressively decreases with time as the
most massive BHs are the first to form binaries
and to be ejected from the cluster.
Thus, in order to relate the evolution of $\mmax$
 to the evolution of $\Mbh$
 we compute the cumulative BH mass distribution (corrected by  ejections due to SN kicks)
 and invert the relation to obtain an expression for the BH mass as a function of
 the current total mass in BHs. Hence, for a generic mass function $\phi$ (after SN kicks, see equation~\ref{eq:phi}) we first solve $\mmax$ from
\begin{align}
\int_{\mlo}^{\mmax}\phi m\dr m = \Mbh,
\label{eq:Mbhint}
\end{align} 
where $\Mbh$ is provided by the cluster evolution (i.e. \model). In some special cases, the solution for $\mmax$ can be written analytically, for example for a power-law $\phi_0\propto m^\alpha$  and no kicks (i.e. $\phi=\phi_0$) we find
 
 \begin{equation}\label{maf0}
\mmax=\left[{\Mbh\over \Mbhn}
\left(\mup^{\alpha+2}-\mlo^{\alpha+2}\right)
+\mlo^{\alpha+2}\right]^{1/(\alpha+2)},
\end{equation}
 where $M_{\rm BH,0}$ is the initial total mass in BHs.
Initially, when $\Mbh= \Mbhn$ we have
$\mmax=\mup$,  while $\mmax=\mlo$ for the last  binary
ejected from the cluster.
Other prescriptions for  $\phi_0$ and $\phi$ could also be  implemented within our theoretical framework to describe
 higher metallicity systems and/or  different 
natal kick prescriptions.

We assume that the BHs receive a natal kick 
given by the momentum conserving model described in Section~\ref{ssec:fretm}.
We then assume  a power-law form for $\phi_0$, i.e. $\phi_0=Am^\alpha$, and for the purpose of finding $\mmax$ we approximate the mass function after kicks by equation~(\ref{eq:phiappr}). {  From integrating this function as in equation~(\ref{eq:Mbhint}) we have
 \begin{align}
\Mbh(\mmax) \simeq 
A\times\begin{cases}
\displaystyle\ln{\left[\left(\qmb^3+1\right)/\left(\qlb^3+1\right)\right]},&\alpha=-2,\vspace{0.2cm}\\
\displaystyle 
\mmax^{\alpha+2}\left(1-h(\qmb) \right)-\mlo^{\alpha+2}\left(1-h(\qlb)\right), &\alpha\ne-2,
\end{cases} 
\label{eq:Mbhhyp}
\end{align}
where $\qml=\mmax/\mlo$, and $\qmb=\mmax/\mb$}.
The constant $A$ is found from solving  in a similar way $\Mbhn\fretm = \int_{\mlo}^{\mup} \phi m\dr m$, which gives the same result as in equation~(\ref{eq:Mbhhyp}), but  with $\mmax$ replaced by $\mup$ (see equation~\ref{eq:fretmbhapprox}). The relation  $\Mbh(\mmax)$ of equation~(\ref{eq:Mbhhyp}) needs to be computed once and can then be inverted numerically to get $\mmax$ for a given $\Mbh$.

Equations~(\ref{maf0}) and (\ref{eq:Mbhhyp}) 
apply only to systems for which 
$\phi_0$ can be described   by a single power-law function.
We now show  that this is a good approximation for clusters with metallicity $Z\lesssim 0.05Z_\odot$.
We obtained an approximation to the  initial BH mass function using the Single Stellar Evolution (SSE) package  \citep{Hurley2002} but with the updated prescriptions for stellar
winds and mass loss in order to replicate the BH
mass distribution of 
\citet{Dominik2013} and \citet{Belczynski2010}.
We sample the masses
of the
stellar  progenitors from a mass function $\phi_{\star}\propto m_{\star}^{-2.35}$  \citep{Kroupa2001}. with  masses in the range
$20$ to $100\,\msun$, and evolve the stars  to BHs 
for the given metallicity.  Fig.~\ref{maf} shows 
that  for metallicities $Z\lesssim 0.05Z_\odot$ the BH
 mass distribution can be fit by a power-low 
 $\phi\propto m^\alpha$ with $\alpha=0.5$ between 
$m_{\rm lo}=3\,\msun$ and $m_{\rm up}=33\,\msun$.
Because the metallicity distribution of GCs in the Milky Way peaks at about $Z\simeq 0.05Z_\odot$ \citep{1985ApJ...293..424Z}, this model can be used to provide a reasonably good
approximation to the  mass function of BHs formed in these
systems.
In what follows we will therefore adopt this model 
 and use equation (\ref{maf0}) or {numerically determined inverse of} equation (\ref{eq:Mbhhyp}) to
describe the evolution of the BH mass function due to
SN and/or dynamical kicks.

\section{Comparison to Monte Carlo simulations}\label{MCcompS}

In this section we show that  \modelb\
reproduces 
the eccentricty distribution, and  rates
of binary BH mergers formed in  detailed  Monte Carlo simulations of  GCs
as well as the evolution of the cluster itself.
Specifically, we compare our  results to the 24 GC Monte Carlo models  of 
\citet{Rodriguez2016a}, \citet{2018PhRvD..98l3005R} and \citet{2018PhRvL.120o1101R}. 
The model initial conditions are given in Table 1 in \citet{Rodriguez2016a};
 these are 24 GC models with initial masses 
in the range  $\simeq 10^5-10^6\,\msun$, and virial radius
$r_{\rm v}=1\,\rm pc$  or $\rm 2\,pc$  (where $r_{\rm h}\simeq 0.8r_{\rm v}$).
For each value of  mass and radius three different 
realisations with metallicity $Z=(0.01,\ 0.05, \ 0.25)Z_\odot$ were evolved
for $\simeq 13\,\rm Gyr$ and during this time
 produced 2819 merging BH binaries, of which 1561 ($0.55$ of the total) were formed inside the cluster, and 
123 ($0.04$ of the total) were GW captures. The rest of the mergers occurred among
the population of ejected binaries.
The Monte Carlo models are self-consistent simulations of the secular evolution of a cluster and include the effect of stellar evolution, a realistic mass function of stars, primordial binaries, the effect of an external tidal field  and employ a three-body integrator (including post-Newtonian terms up to order 2.5) in order to accurately follow the strong binary-single interactions which lead to the hardening and merger of the core binaries. 

\begin{figure}
\centering
\includegraphics[width=3.in,angle=0.]{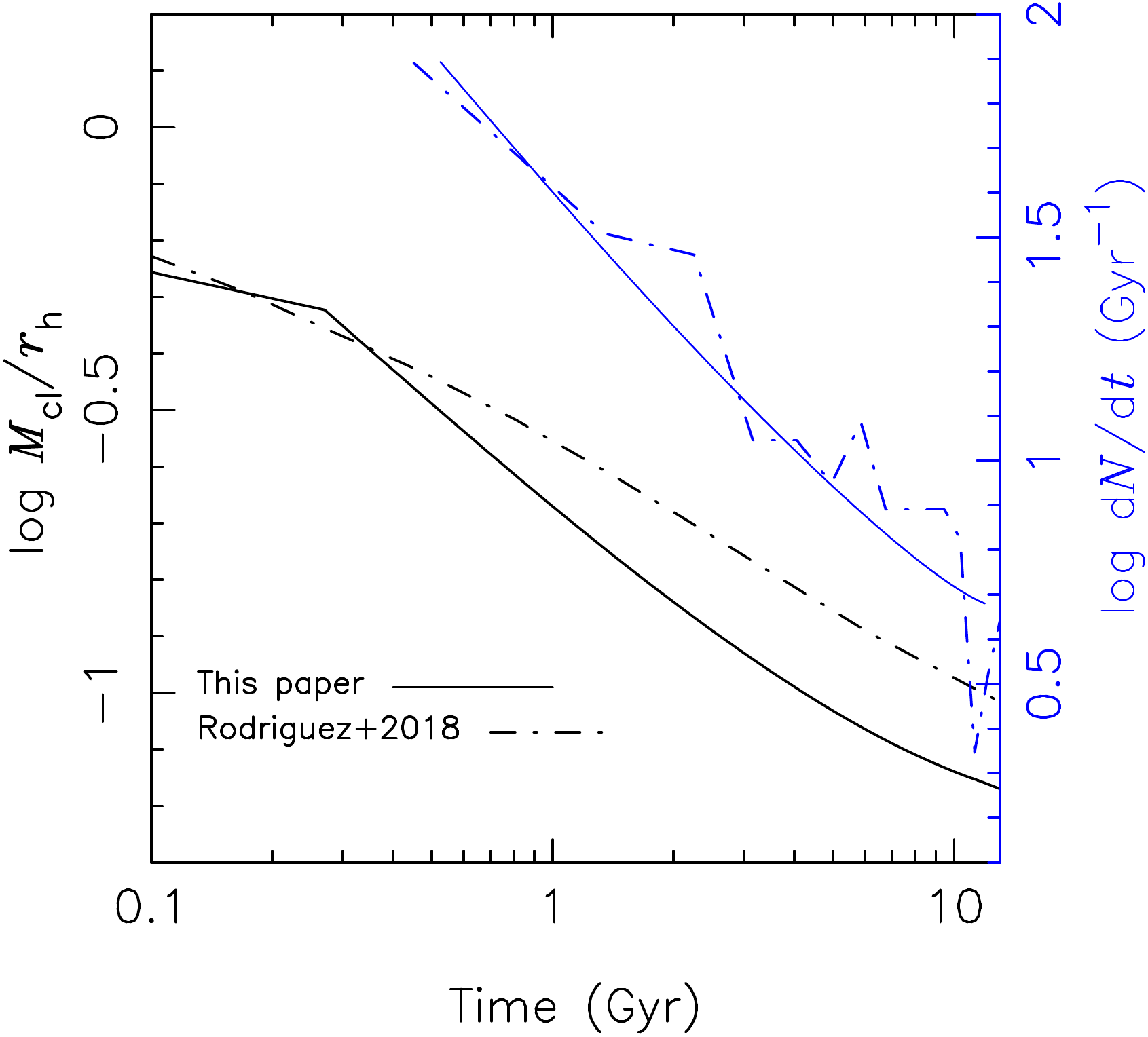}
\caption{Evolution of the cluster compactness, that we define  as
$M_{\rm cl}/r_{\rm h}$, and the BH binary merger rate predicted by
our method. We  compare this to a similar Monte Carlo model from  \citet{2018PhRvD..98l3005R}. The cluster mass and
radius have been normalised to their initial values $M_{\rm cl,0}=1.3\times10^6\,\msun$
and $r_{\rm h,0}=0.8\rm pc$.
}\label{Expand}
\end{figure}

For each value of cluster half-mass radius and total mass, we  used equations~(\ref{eqN}) and (\ref{eqNe}) to compute
the rate  and eccentricity distribution of binary BH mergers that occur after a given time.
Specifically we consider mergers that occur after $3$, $5$ and $8$ Gyr  from the start of the simulation;
if we assume that the clusters form at $z_0=3$,
then, for the cosmological parameters above, these times correspond to binaries that merge
within a redshift $z_{\rm d}\simeq 2,\ 1$ and $0.5$  respectively.  
The distributions from all the cluster models 
were then added together and compared to those
from all the  binary BH mergers produced in the Monte Carlo
simulations.  We only show the comparison for  models
where the BHs received   the same momentum kick as neutron stars (with $\sigmans=265~\kms$).
We note, however, that because  of the high retention fractions in these  systems, models  without kicks applied to the BHs produce similar eccentricity distributions
as models with kicks applied.  
Moreover, we use here an initial BH mass fraction $f_{\rm BH}=0.04$,
appropriate for the low metallicity clusters considered.

Fig.~\ref{mccomp} shows the eccentricity distributions 
at the moment the binaries first achieve  a GW frequency of
$10\,\rm Hz$ (i.e., near the low end of the {LIGO-Virgo} frequency window).
The two methods give a very similar number of mergers 
 at any given $e$
 over the entire range of  eccentricities (i.e., $e>10^{-7}$), and,
  for example, both produce $\sim 5\%$ of BH mergers with $e>0.1$ inside the LIGO-Virgo band.
   The individual eccentricity distributions of in-cluster mergers and mergers among the ejected binaries shown in the bottom panels
 also overlap reasonably well with those from the Monte Carlo simulations
 as do the fraction of mergers produced
by each of the three channels  (upper panels of Fig. \ref{mccomp2}),
and  the corresponding
total number of mergers  per cluster at a given time  (lower panels of Fig. \ref{mccomp2})
for the entire cluster mass range considered.

In view of the approximate nature of our method, the agreement with 
the Monte Carlo results can appear quite remarkable. However,
it is not a coincidence, but  a rather natural consequence of H\'{e}non's principle.
This principle, which is the basis of
our approach,  is also  what determines the relation between the evolution of a
cluster Monte Carlo model and the merger rate and 
properties  of the BH binaries that are produced in its core. 
To illustrate this latter point, 
we plot in
Fig. \ref{Expand} the  evolution of the global cluster `compactness', defined here as $M_{\rm cl}/r_{\rm h}$, 
 and the corresponding merger rate of BHs for
 a system with initial mass and radius
$M_{\rm cl,0}=1.3\times10^6\,\msun$, $r_{\rm h,0}=0.8\,\rm pc$. We compare our results to those from a Monte Carlo cluster model
with similar initial mass and radius \citep[from Fig.~4 of][]{Rodriguez2016a}.
In both cases,  two-body relaxation causes the cluster radius 
to expand significantly over the 
timescale of the simulation, leading to a decrease in the BH binary merger rate.
The physical reason why the merger rate goes down is because the cluster energy demand decreases as the cluster expands
and a significant drop in the binary hardening rate happens after about a relaxation time.

\begin{figure*}
\centering
\includegraphics[width=7.in,angle=0.]{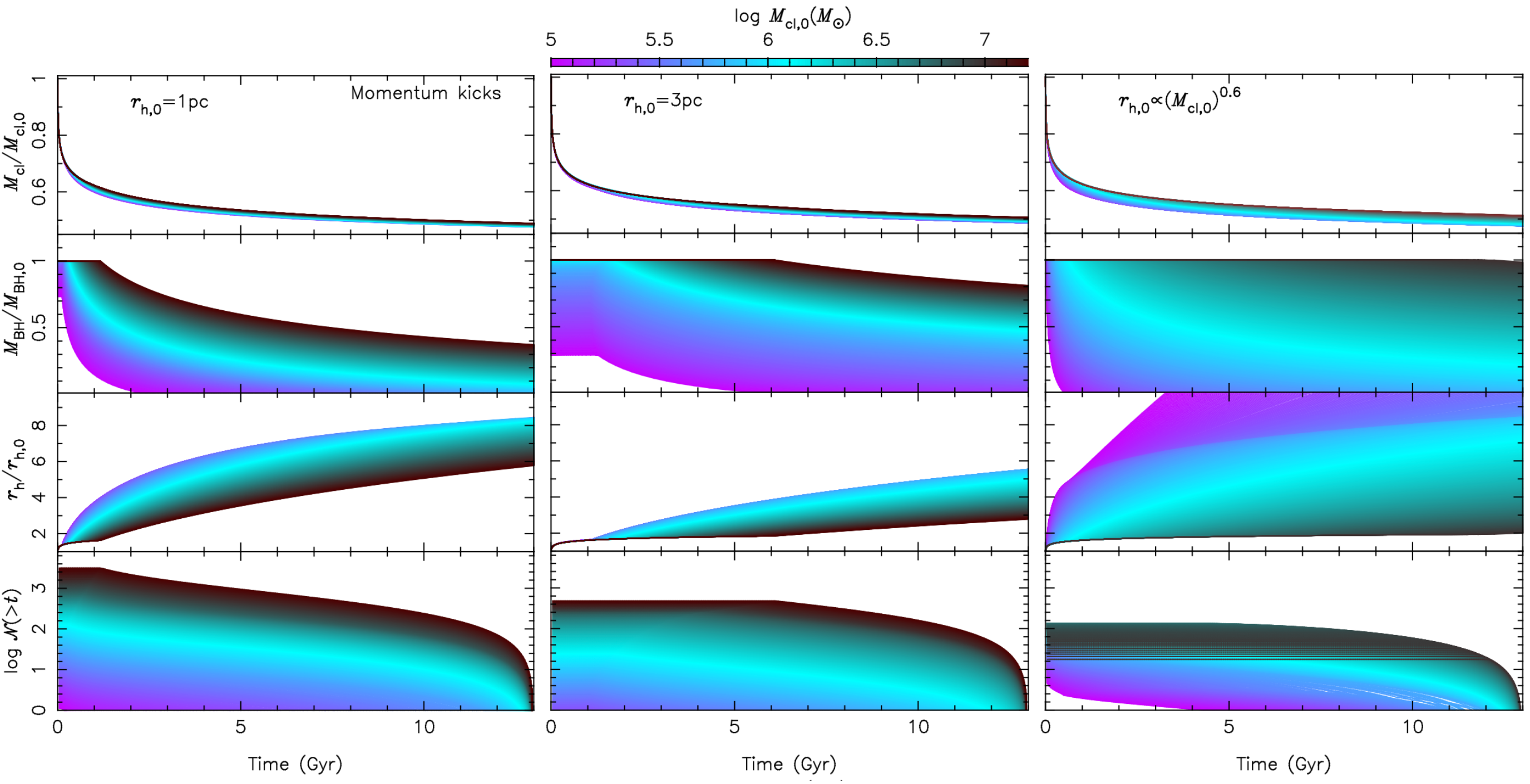} \caption{
Evolution of cluster models for a range of masses
and initial cluster radii. From top to bottom the panels  show the evolution of the total mass of the clusters, the evolution of the mass  in BHs, the cluster radius, and the
number of  BH binaries that merge 
 after a given time  from the start of the simulation. In this case we have adopted
the momentum conserving natal kick model described in Section\ \ref{ssec:fretm}.
}\label{evol1}
\end{figure*}
\begin{figure*}
\centering
\includegraphics[width=7.in,angle=0.]{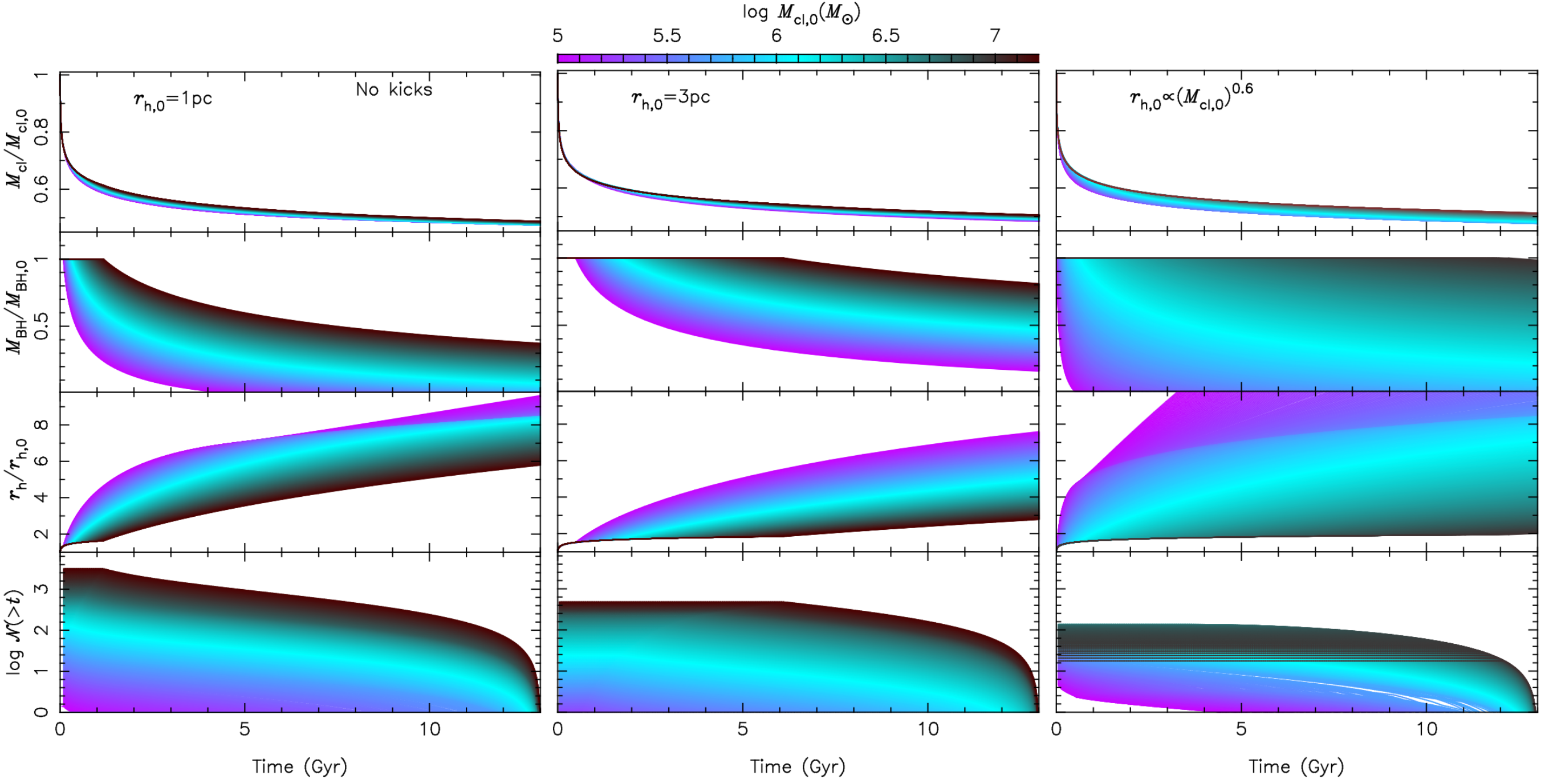}
\caption{
Same as Fig. \ref{evol1} but for models where
the BHs receive no natal kicks.
}\label{evol2}
\end{figure*}

\section{Astrophysical implications}\label{AI}
In this section we use  \modelb\ to derive the 
number and
eccentricity distributions of merging BH binaries
and study their relation to the initial properties of the parent cluster.
Whilst a more detailed analysis will be presented elsewhere,
here we simply consider how such distributions 
are linked to  the initial mass and radius of a GC and  its time evolution.
The new method
 allows us to explore for the first time a wide range
 and physically motivated set of initial conditions,
 spanning more than two orders of magnitude both in radius and mass.

\subsection{Cluster evolution and merger rates}
We consider  the evolution of a set of models with initial  mass 
in the range $M_{\rm cl,0}\simeq(10^5,\ 10^7)\,\msun$ that we evolved for 13\,Gyrs. We adopt three different models for the initial radius of the clusters. We either assume that the initial radius is independent 
of mass and set  $r_{\rm h,0}=1\,\rm pc$
or $3\,\rm pc$, or we assume an  initial mass-radius relation 
\citep{2010MNRAS.408L..16G}:
\begin{equation}\label{FJ}
\log \left(r_{\rm h,0}\over {\rm pc} \right)=-3.560+0.615\log \left(M_{\rm cl,0}\over \,\msun \right) \ .
\end{equation}
This latter relation was obtained by converting the original Faber-Jackson relation of elliptical galaxies 
to a mass-radius relation \citep{2005ApJ...627..203H}, and correcting for the adiabatic expansion due to mass-loss caused by stellar evolution   \citep{2010MNRAS.408L..16G}.
We show the results for models with constant momentum  natal kicks and no natal kicks  in Fig.~\ref{evol1} and Fig. \ref{evol2},
respectively. We take hereafter an initial BH mass fraction of $f_{\rm BH}=0.04$.

The top panels in Fig.~\ref{evol1} and Fig. \ref{evol2} show that the total cluster mass has decreased by approximately a factor of two by the end of the simulation  and that most of this mass-loss occurs during the first few Gyrs of evolution.
Equation (\ref{eq:Mstdotsev}) implies that the fractional stellar mass-loss per unit time is the same for all clusters, so the small differences
seen in the evolution  of $M_{\rm cl}$  are a consequence of the dependence of $\dot{M}_{\rm BH}$ on the cluster relaxation time, and therefore on its mass and radius. 
The mass-loss from the system due
to stellar evolution also causes the cluster radius to expand by approximately a factor 2. This
can
be seen in the figures  at early times before the 
subsequent expansion due to two-body relaxation starts to dominate.

The top-middle panels in Fig.\ref{evol1} and Fig. \ref{evol2} show that the evolution of the  BH population 
is quite sensitive to the  initial cluster properties. Systems that are initially more compact and with a low mass have a shorter relaxation time, and
consequently they start  to process their BH binaries  at earlier times (i.e., at $t_{\rm cc}$) and at a higher rate than larger and more massive clusters (see equation\ \ref{eq:mbhej}).
{  For example, the $r_{\rm h,0}\propto M_{\rm cl,0}^{0.6}$  models are highly dynamically active for $M_{\rm cl,0} \lesssim 10^5M_\odot$, yielding rate of cluster expansion, BH depletion, and number of merging binaries  of larger magnitude than their $r_{\rm h,0} =1~$pc and 3~pc counterparts \footnote{Interestingly, for
the $\sim10^5M_\odot$ clusters, the extent of the initial expansion is comparable to that due to a residual gas expulsion (assuming an overall, uniform star formation efficiency of $30\%-40\%$) from a similarly compact proto-cluster \citep{2017A&A...597A..28B}.}. }

\begin{figure}
\centering
\includegraphics[width=3.in,angle=0.]{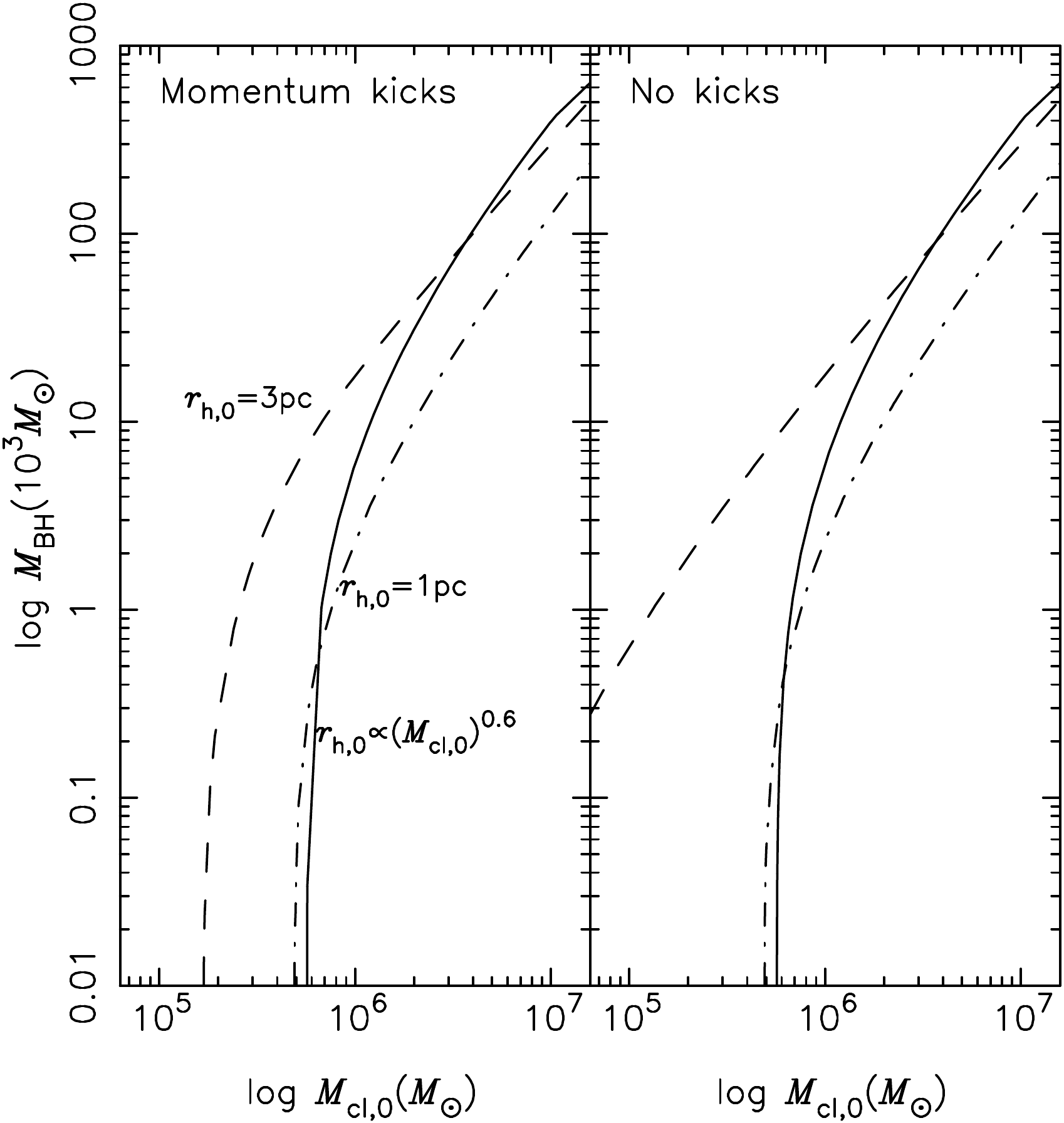} 
\caption{Mass in BHs left after 13 Gyrs of evolution as a function of initial cluster mass for the systems
of Fig. \ref{evol1}
and Fig. \ref{evol2}.}\label{mvsm}
\end{figure}

As a result of their long half-mass relaxation time, all cluster models with $M_{\rm cl,0}\gtrsim 10^6\,\msun$ still have BHs after 13\,Gyrs of evolution for the initial radii we adopted, {  corroborating the recent inference of stellar-mass BHs in $\omega$Cen \citep{Zocchi2019, 2019MNRAS.488.5340B} and 47~Tuc  \citep{2019arXiv190808538H}.}
Clusters with mass lower than this can also  retain some of their black holes if their densities are sufficiently low initially (e.g.,  $r_{\rm h,0}=3\,\rm pc$).
These results are also illustrated in Fig. \ref{mvsm} which
gives the total mass in BHs at $t=13\,$Gyr for all the models
considered.
This figure  demonstrates that the final 
number of BHs  in  clusters with  
 $M_{\rm cl,0}\gtrsim 5\times 10^5\,\msun$
does not depend significantly  on the assumptions we make about the BH
natal kicks. But, models with an initial mass lower than this value  retain a significant fraction
 of their BHs only if no kicks are applied and the cluster initial radius is large, $r_{\rm h,0}=3\,\rm pc$. 
We conclude that whether a cluster will be able to retain some of its BHs  depends on two main factors:
 the half-mass relaxation time of the whole system, which determines the rate at which the BHs are ejected
from the cluster, and  the initial BH fraction which is
set by the stellar initial mass function, metallicity, and most importantly, by the  natal kicks.
{  These results are consistent with recent work based on Monte Carlo simulations
\citep{2018MNRAS.478.1844A,2019ApJ...871...38K}, while extending  these conclusions
to a much wider region of relevant parameter space for star clusters.}

After $\tcc$, a system reaches  balanced evolution, and its half-mass radius
 starts to increase as the result of  two-body relaxation.
The evolution of $\rh$ for our models
appears to be very sensitive to the initial conditions.
Generally, systems that have a shorter half-mass relaxation time expand faster 
and end up with a larger $\rh$ after a Hubble time. 
The evolution of the cluster radius should also depend on metallicity,
the initial stellar mass function, and the assumed natal kick distribution because these 
 set  the initial fraction of BHs,
which changes  $\trh$ through the parameter $\psi$.
The effect of natal kicks becomes especially important 
in systems with a low mass  ($M_{\rm cl,0}\lesssim 5\times 10^5\,\msun$) and a large half-mass radius ($r_{\rm h,0}\gtrsim 3\,\rm pc$) due to the larger fraction of  BHs  that are ejected.
These results fit in the view that  the 
 properties of
   the whole system determine the evolution of the BH sub-system \citep{2013MNRAS.432.2779B}, but  the evolution of the
   cluster itself is also affected by the remaining mass fraction in BHs (Section \ref{ssec:coevolv}).

\begin{figure*}
\centering
\includegraphics[width=3.5in,angle=270.]{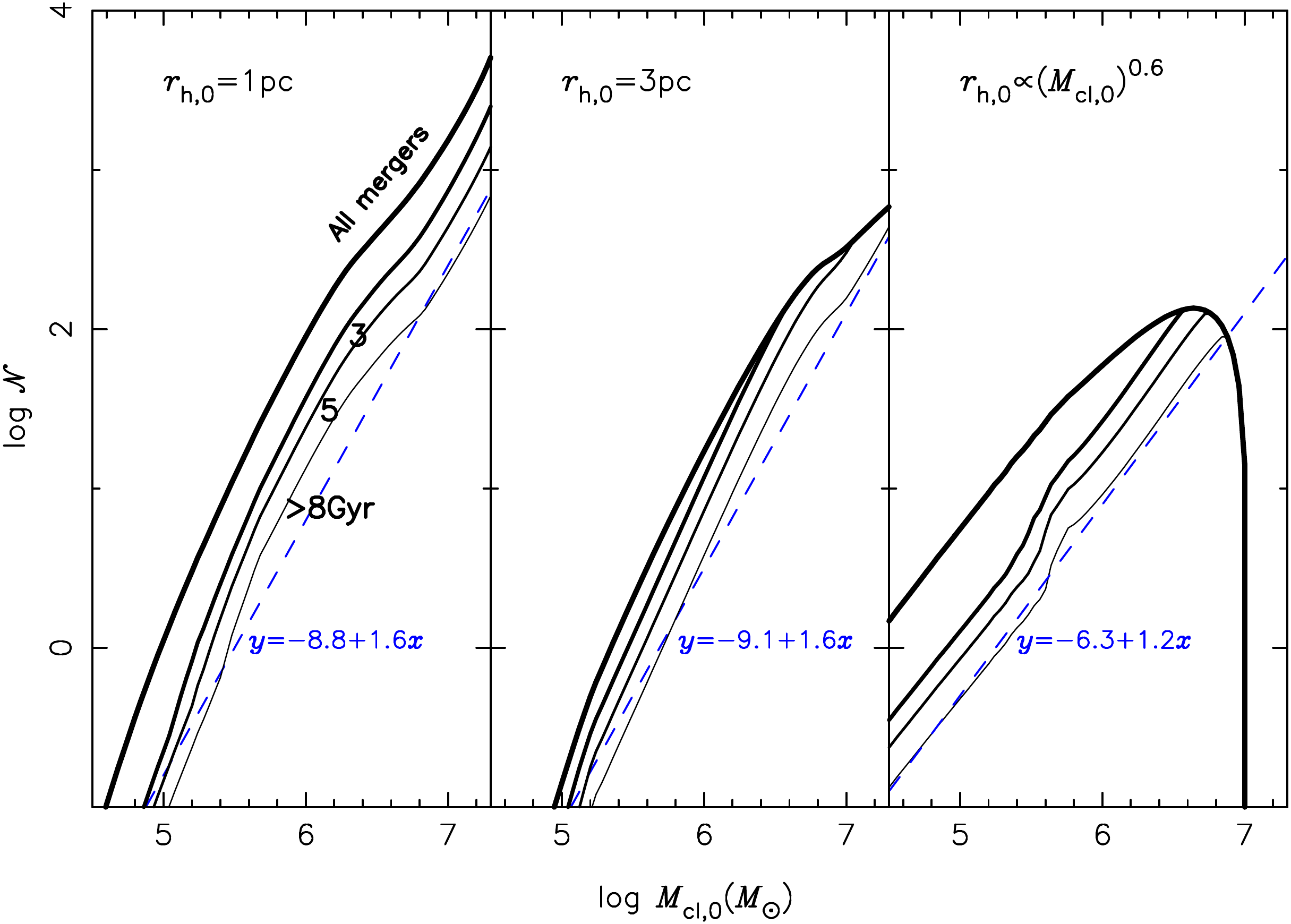} 
\caption{Number of merging binaries as a function of cluster mass  for the models of Fig. \ref{evol1}
where the BHs receive  the same momentum kick as neutron stars.
From top to bottom line, we show the number of  binaries which merge
at times $t>0$, 3, 5, and 8\,Gyr
from the start of the simulation.
Blue lines show the best  fit models to the mergers that occur at late times.
}\label{nvsm}
\end{figure*}

The total number of BH binaries that merge at
times $>t$ can be
seen in the bottom panels of Fig.~\ref{evol1} and Fig. \ref{evol2}. Clusters with  a longer relaxation time  start forming binary BHs later,
which explains the initial lag  between the start of the simulation and when the first
merger occurs (i.e., the plateau of $\mathcal{N}(>t)$ seen at early times). 
The BH binary merger rate is shown as a function of cluster  properties  in 
Fig.\ \ref{nvsm}.
 Assuming a constant initial radius, the total
number of BH binaries that merge at  late times ($t>8\,$Gyr)
 can be reasonably well fit by a simple  scaling
$\mathcal{N}\propto M_{\rm cl,0}^{1.6}$. 
For the relation equation\ (\ref{FJ}), instead,
 the best  fit model is $\mathcal{N}\propto M_{\rm cl,0}^{1.2}$,
 for $M_{\rm cl,0}\lesssim 10^7\,\msun$,
 {  while for more massive clusters the  segregation timescale of the BHs   exceeds the Hubble time and no  mergers are produced in these systems. }
 Taken together, our results imply 
 that $\mathcal{N}\propto M_{\rm cl,0}^{1.6}r_{\rm h,0}^{-2/3}$.
  These results can be compared to  \citet{2018MNRAS.480.5645H} who
  find that the total number of mergers produced by their cluster models over 12\,Gyr  scaled  as $\mathcal{N}\propto M_{\rm cl,0}^{1.3}r_{\rm h,0}^{-0.9}$. 
  The difference with our best fit model could be due to the larger
 parameter space explored by us and by the fact that 
 these authors did not include
in-cluster mergers in  their Monte Carlo simulations. 

\begin{figure*}
\centering
 \includegraphics[width=6.8in,angle=0.]{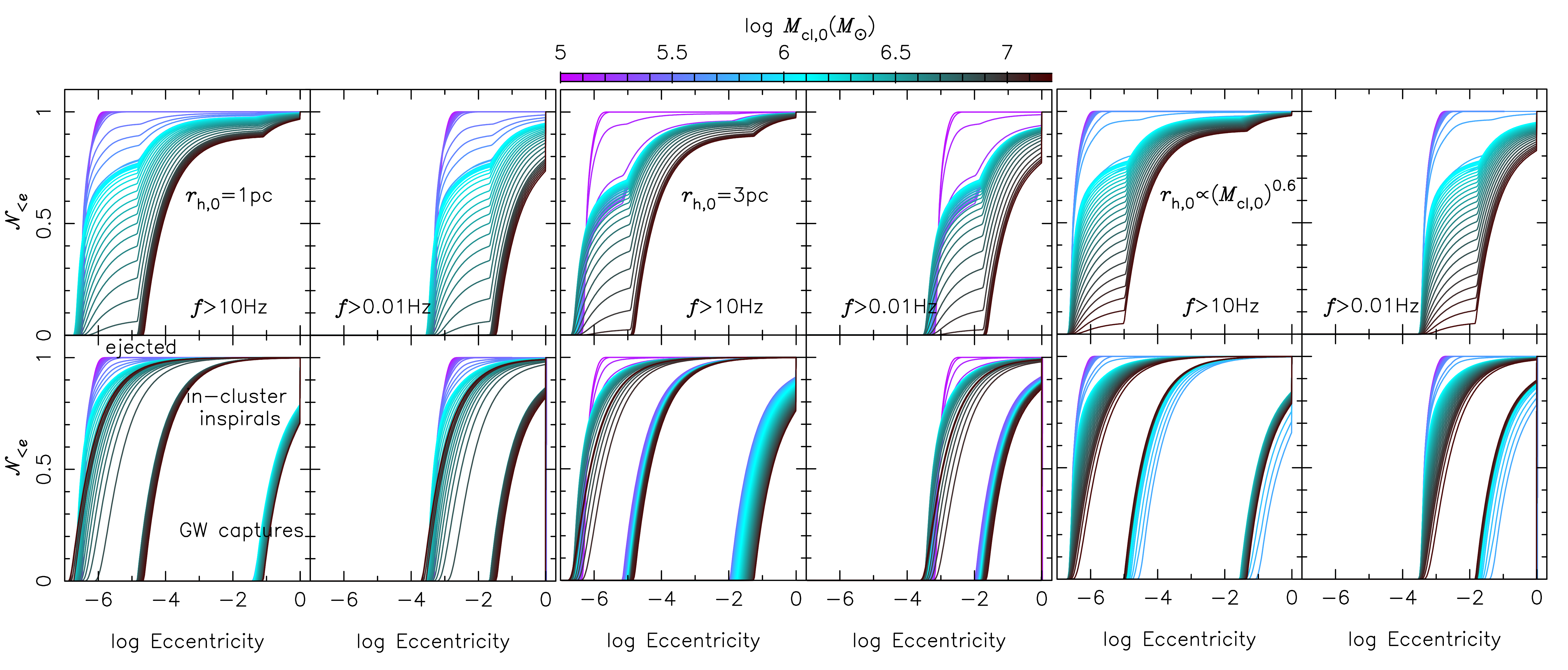}
\caption{Eccentricity distributions of merging BH binaries formed in star clusters for the three choices of $r_{\rm h,0}$, at the moment
their GW frequency becomes larger than $10$ and $0.01$\,Hz. Bottom panels give the eccentricity distributions for  ejected binaries (lowest eccentricities), in-cluster inspirals and GW captures (largest eccentricities) separately. Note that at $0.01$\,Hz, GW captures 
all have $e\simeq 1$ because thy  all form at frequencies $f>0.1\,$Hz. In this calculation we have adopted the constant momentum natal kick model of Section \ref{ssec:fretm}, and we have  only included late time mergers, occurring  at $t>8$\,Gyr.
}\label{edis}
\end{figure*}
As a general trend, we observe a strong dependence of the merger rate on the cluster initial 
conditions, with
 a larger cluster mass and a smaller radius
resulting into an overall higher BH merger rate. 
These results
put into question
some of the current literature
where  simplified assumptions about the initial 
 mass/half-mass radius relation were adopted. For example,
\citet{2018ApJ...866L...5R} derived their binary BH merger rate 
assuming that $50\%$ of clusters form with a virial  radius of
$r_{\rm v}=1\,$pc and $50\%$ form with $r_{\rm v}=2\,$pc.
Similarly, \citet{2018PhRvL.121p1103F} assumed that the 
binary BH merger rate is independent of the cluster radius for which they assumed a fixed
value. \citet{2017MNRAS.464L..36A} derived a merger rate by using a limited set of models
with three values of cluster initial radii and four values of cluster mass.
These restrictions were
due to time constraints imposed by the computationally  expensive techniques employed in these studies.
Although our models include less physics, the new treatment allows for a complete 
exploration 
of the parameter space relevant for GCs. This
opens the possibility for a first realistic determination of
the merger rate and properties of BH binaries produced in star clusters which we reserve to
a future paper. We note that
a  similar attempt was made before in \citet{2019ApJ...873..100C} who adopted the
semi-analytical  approach  of \citet{2016ApJ...831..187A} to estimate a cosmic black hole merger rate from GCs.
In \citet{2016ApJ...831..187A}, however, the binary hardening rate was computed from the cluster  core density which
 cannot be easily linked to the secular evolution of a cluster. Here, we have 
overcome this issue by using
 H\'{e}non's principle to relate  the binary hardening rate to the evolving 
 \emph{global} properties of its host cluster.

\begin{figure*}
\centering
\includegraphics[width=2.6in,angle=270.]{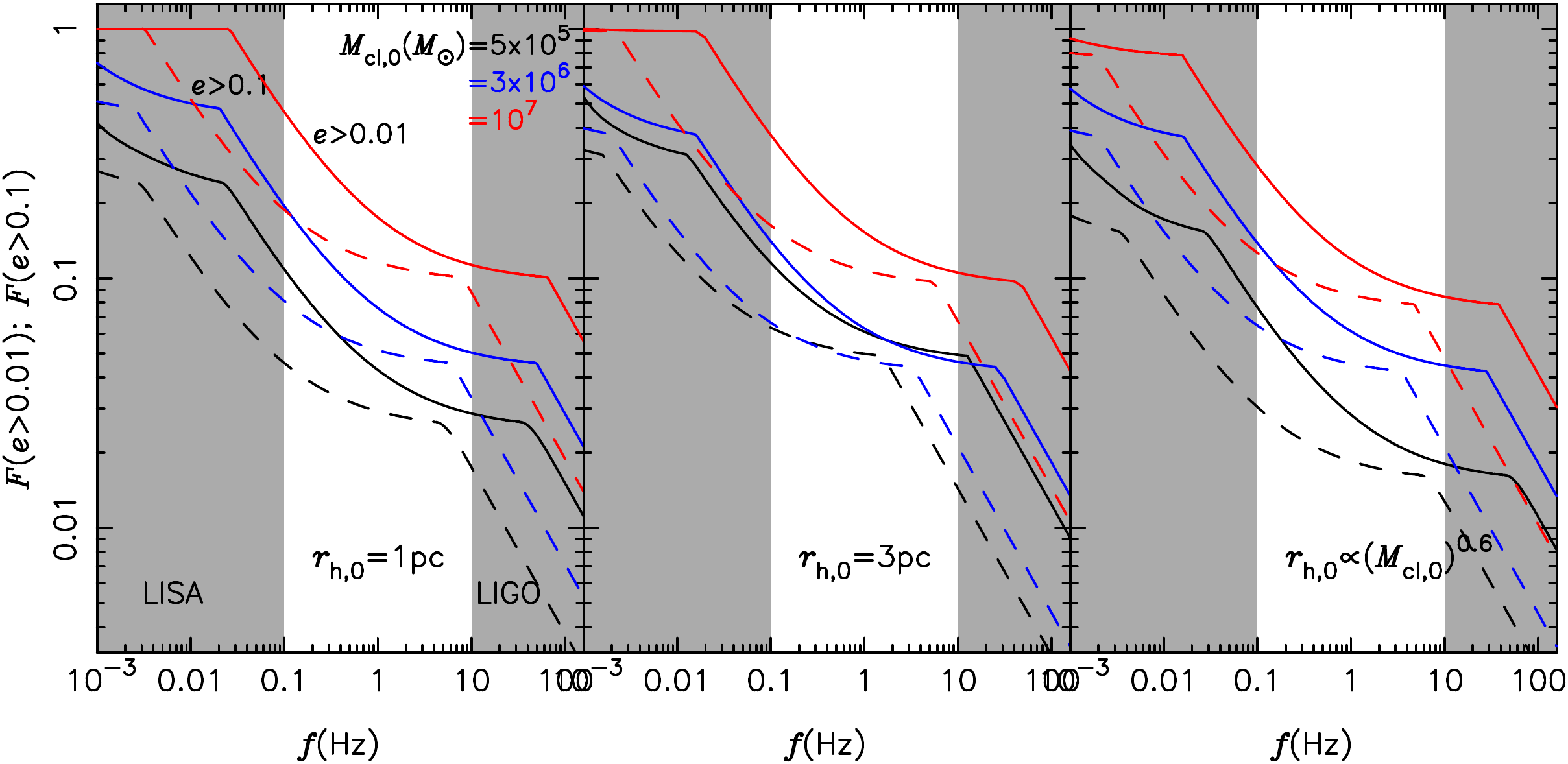} 
\caption{Fraction of binaries that have an eccentricity larger than 
$0.1$ and $0.01$ as a function of peak GW frequency. 
We have considered three representative values of  cluster mass 
for our three choices of radius,
 have adopted the natal kick model of Section \ref{ssec:fretm}, and only 
  binaries that merge after a time larger  than 8\,Gyr
from the start of the simulation have been included in the analysis.
}\label{ecc2}
\end{figure*}

\begin{figure*}
\centering
\includegraphics[width=3.5in,angle=270.]{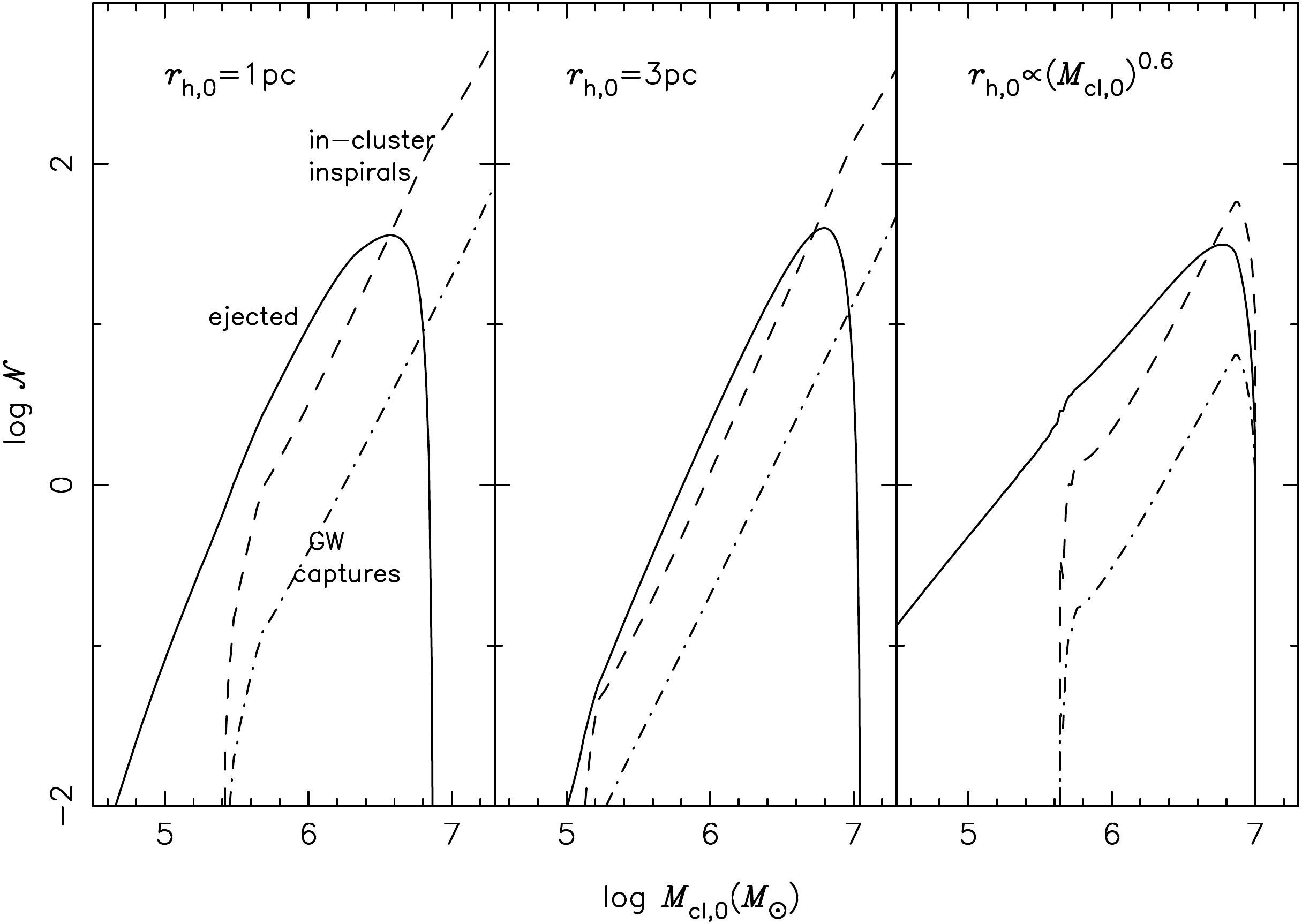} 
\caption{Number of  BH binaries that  merge  after  $8$\,Gyr
as a function of cluster mass,  for the models of Fig. \ref{evol1}
and for the three merger channels shown separately.
}\label{nvsmD}
\end{figure*}

\subsection{Eccentricity distributions}\label{ediss}
Binary black holes formed in dynamical environments, such as globular clusters, 
and nuclear star clusters, can have a finite  eccentricity in both {LIGO-Virgo}, and the LISA bands \citep[e.g.,][]{2016PhRvD..94f4020N}. 
For  mergers formed through the evolution of field binaries instead,
any  eccentricity is washed out  by GW radiation by the time the signal enters 
the detector frequency band \citep{2002ApJ...572..407B}. As a non-zero eccentricity would be a unique fingerprint of a dynamical origin, we focus in this section on the relation between 
a cluster properties and the
eccentricity distribution of the BH binary mergers that it produces.

In Fig. \ref{edis} we show the eccentricity distribution of  the merging  BH binaries.
We focus here on mergers occurring at late times ($t>$8\,Gyr) because these are the most relevant for current and planned GW detectors. If we assume that the clusters form at $z_0=3$,
then, for the cosmological parameters above, this time  corresponds to binaries that merge
within a redshift $z_{\rm d}=0.5$. 
We show these distributions  for several values of $M_{\rm cl}$, for our three  choices of cluster
radius, and at
the moment the binary GW frequency  first becomes
larger than   $10$ and $0.01$ Hz. 
We chose to plot the eccentricity distributions 
at $10$ and $0.01$ Hz because these are near the
lower end of the LIGO-Virgo and LISA frequency windows respectively \citep{2010CQGra..27h4006H,2017arXiv170200786A}. 
In the bottom panels the cumulative eccentricity distributions 
of in-cluster inspirals, GW captures, and mergers among the ejected binaries
are shown separately.
 
Our analysis demonstrates  that independently of  the cluster initial conditions, GW captures have   $e>10^{-2}$ at $f>10\,\rm Hz$, while 
all other type of mergers have eccentricities at these frequencies ($e\lesssim 10^{-2}$) that might be too small to be measurable using current detectors \citep[e.g.,][]{2013PhRvD..87l7501H,2018PhRvD..98h3028L,2019ApJ...871..178G}.  
 As expected, mergers among the ejected binaries have the smallest eccentricities.
We conclude that GW captures are the only type of mergers that we have considered in our analysis which  could have a measurable residual eccentricity at the moment they 
enter the LIGO-Virgo  frequency window. 
At $f>10\,\rm Hz$, in-cluster mergers show a clear tri-modality, with the lower peak corresponding to isolated binaries that merge after a dynamical encounter,  the higher eccentricity population  ($0.1<e < 1$) corresponding to sources which merge during an encounter via a GW capture, and a third population of mergers with $e\simeq 1$. These latter systems also merge through a  GW capture, but form directly at frequencies larger than $10\rm Hz$ rather than reaching this frequency  after substantial circularisation has already occurred.

Because all GW captures are formed at frequencies
 above $\simeq 0.1\rm Hz$,  
 the eccentricity distributions of these mergers appears as a delta function  at  $e= 1$ for the lowest frequency range considered in Fig. \ref{edis}.     As shown in \citet{2017ApJ...842L...2C}, these highly eccentric sources might
 elude detection  because the high eccentricity shifts the peak of the relative power of the GW harmonics towards higher frequencies,
so that their maximum power is emitted  farther away from the  frequency band 
of interest.
Mergers formed by GW captures are therefore unlikely to be detectable at frequencies lower than $0.1\rm Hz$.
On the other hand, 
for $M_{\rm cl, 0}\gtrsim 10^6\,\msun$ we see that all in-cluster inspirals reach  $f=0.01\rm Hz$
 with $e\gtrsim 0.01$, which is potentially measurable with planned  missions such as LISA \citep{2016PhRvD..94f4020N} and third generation detectors \citep{2010CQGra..27h4007P}.

 The fraction of eccentric GW sources  is shown in Fig.~\ref{ecc2} 
as a function of the peak frequency $f$ and for three representative values of cluster mass. From this plot 
   we see that  between $\approx 30$ to $100\%$ of our binaries  start eccentric inside 
   the LISA frequency window ($0.001-0.1\rm Hz$),
and many  retain a significant eccentricity by the time they 
reach $f\simeq 0.01\rm Hz$ where LISA is most sensitive.
 The  exact fraction of  systems that start eccentric in the LISA band  
 depends on the cluster  properties, 
 reaching unity for $M_{\rm cl,0}\simeq 10^7\,\msun$.
   This is different from BH binary mergers formed from the evolution of field binaries
   of which only $\lesssim10\%$ are expected to have $e>0.01$ at 
   $f=0.001\rm Hz$ \citep{2016ApJ...830L..18B}. 
   We conclude that sufficient observations of the eccentricity of binaries in the LISA band
   could be used to  discriminate between formation in the field  and in clusters \cite[e.g.,][]{2016PhRvD..94f4020N},   
  but  could also help to determine
  which type of clusters are most likely to produce the  binaries.
 
 The overall  eccentricity distribution of the BH binaries is most sensitive to the relative fraction
 of the three type of mergers. We show
 how these fractions depend on the cluster  properties  in Fig.\ \ref{nvsmD}.
 Three regimes can be identified: (i)~if  $a_{\rm GW}>a_{\rm ej}$
 all binary black hole mergers are produced inside the cluster.
 About $\lesssim10\%$ 
of these in-cluster mergers are GW captures and the rest are mergers
that occur in between  binary-single encounters. 
 For the systems in Fig.\ \ref{nvsmD}  this regime can be seen
at $M_{\rm cl, 0}\gtrsim 10^7\,\msun$;
(ii)~when $a_{\rm GW}<a_{\rm ej}$ almost half of the mergers occur inside the cluster,
 half are  mergers  among the ejected binaries, and
less than a few per cent of the mergers are GW captures.
For the clusters in  Fig.\ \ref{nvsmD} this  regime
occurs at $10^6\,\msun \lesssim M_{\rm cl, 0}\lesssim 10^7\,\msun$;
(ii)~if the BH binary population has
been fully depleted by a time $t$, then all  mergers observed at 
later times will come from the ejected binary population.
This third regime occurs for the lowest mass systems in Fig.\ \ref{nvsmD}
where we see that the  number of in-cluster mergers goes to zero.
The exact transition between the three regimes  depends on the initial mass-radius relation and the time between the formation of the cluster
and when the mergers are detected. 

Our results show that for a typical GC the fraction of in-cluster mergers 
can be quite large, in  agreement 
with  recent studies showing 
that nearly half of all binary black hole mergers occur inside the cluster
\citep{2018PhRvD..97j3014S,2018PhRvD..98l3005R}.  On the other hand, our models also show  that the number  of in-cluster mergers 
detectable by LIGO-Virgo
will be quite sensitive to the uncertain
initial cluster mass-radius relation and
initial mass distribution of the clusters.
 Notably, in the most massive GCs
($M_{\rm cl,0}\gtrsim 10^6\,\msun$)
we would expect a large fraction ($\gtrsim 0.5$) of mergers to be produced inside the cluster rather than 
 happening among the ejected binaries.
   In even higher velocity dispersion clusters such as nuclear clusters, all mergers  should typically 
be formed while the BH binaries are still bound to the cluster, although a
significant contribution from the ejected population might be expected 
near the low end of the nuclear cluster mass distribution \citep{2016ApJ...831..187A}.

{  We note, finally,  that  detailed post-Newtonian direct $N$-body simulations show that in low mass clusters ($M_{\rm cl}\lesssim 10^5M_\odot$), in-cluster mergers are primarily triggered by long-term  interactions involving stable and marginally stable triples
\citep{2014MNRAS.441.3703Z,2016MNRAS.463.2443K,2017MNRAS.467..524B,2018MNRAS.473..909B,2018MNRAS.481.5123B,2019MNRAS.483.1233R}.  A large fraction of these mergers will have a finite eccentricity in the LIGO band but are not included in our analysis.}

\subsection{Caveats and discussion}
In deriving the merger probability and eccentricity distributions above we have made a few important  
assumptions which we now discuss and justify.

We have assumed that the dynamical interactions only occur between BHs. This is reasonable because 
due to mass segregation  the BH densities  near the core are expected to be much larger than the densities of stars. BHs will therefore dominate the interactions. 
Moreover, because exchange interactions tend to pair the BHs with the highest mass and tens of three body encounters are required before a merger, then mergers will be primarily between BHs when they are present \citep[e.g.,][]{Sigurdsson1993,2015ApJ...800....9M}. As the number of BHs in the core decreases,   interactions with main sequence stars, giant, and white dwarfs
can become more frequent. These interactions could lead to the formation of mass-transferring binaries
with an observable electromagnetic signature \citep[e.g.,][]{1975MNRAS.172P..15F,2010ApJ...717..948I}, {  or detached BH-stellar binaries that can be identified with radial velocities \citep{2018MNRAS.475L..15G}}.  {  Note also that the removal of BHs strictly from the upper end of its mass distribution and the interactions among the most massive BH members, as assumed here, is an idealisation. Detailed $N$-body computations suggest that dynamically assembled binaries span over a mass ratio between 
$0.5-1$  \citep{2017MNRAS.467..524B,2018MNRAS.481.5123B}, implying that BHs over a considerable mass range could be involved in mutual encounters.}

The eccentricity distributions were derived by assuming that the  binaries 
have a large eccentricity, $e_0\approx1$ , at the moment their evolution starts to be dominated by GW radiation.
This allowed us to find the simple analytical relation  equation (\ref{eapp})  between
the eccentricity of a binary and its peak GW frequency.
The assumption of high initial eccentricity  is reasonable for in-cluster mergers, 
but might not  be  valid for some fraction of the  mergers occurring among the ejected binaries \citep{2019PhRvD..99f3003K}. 
The eccentricity distribution of all ejected binaries  is thermal, but for
 the subset of these binaries that  merge within one Hubble time 
the distribution is skewed towards higher
eccentricities.
Thus, we expect  our approximation to hold
as long as $p_{\rm ex}(a_{\rm ej})\ll 1$, and
to  progressively worsen as $p_{\rm ex}(a_{\rm ej})$  increases.
By setting $p_{\rm ex}(a_{\rm ej})=1$ and
solving for $v_{\rm esc}$
we find  that for escape velocities larger than
\begin{align}\label{emer}
{v}_{\rm ex}\simeq 19\,{\rm km\ s^{-1}}\left({m_1m_2\over m_{123}}q_3{6\over {\msun}}\right)^{1/2}\left({{\rm10\,Gyr}\over 
\tau}\right)^{1/8}\ ,
\end{align}
all ejected binaries will merge and their eccentricity distribution will  be thermal.
Even in this case, however,  the overall impact on the 
eccentricity distribution  should not be great as only about $10\%$  of the ejected binaries  
 have $e_0<0.3$. Moreover, the contribution of the ejected binaries becomes less important 
for the more massive clusters that satisfy  the condition $v_{\rm esc}>\tilde{v}_{\rm ex}$ (see Section~\ref{ediss}).

We note also that our calculation is based on the assumption that most binary BH mergers
are formed through strong binary-single interactions. However,
BH mergers in star clusters can also occur  through other processes which  have not be
considered in our analysis. These include mergers mediated by the Lidov-Kozai mechanism
in hierarchical triples  \citep{2002ApJ...576..894M,Wen2003,Kimpson2016}, non-hierarchical 
triples \citep{2014ApJ...781...45A,2016ApJ...816...65A,2018arXiv180506458A}, 
 mergers from direct BH-BH captures \citep{1990ApJ...356..483Q,Kocsis2012,2019arXiv190203242R},   and
eccentric mergers during binary-binary strong interactions \citep{2019ApJ...871...91Z}.
All these effects are  believed to  play only a marginal role (at a $\sim 1\%$ level)  
for the BH merger rate, but might somewhat increase  the number of  binaries
that enter the LIGO-Virgo frequency band with a large eccentricity. 
More recently,
\citet{2019arXiv190409624H}
and \citet{2019arXiv190607189S} argued that weak fly-by encounters could also
affect the eccentricity distributions of in-cluster mergers.

{  Finally, primordial binaries which are not considered here are an essential ingredient of the early dynamical evolution of globular clusters and their young progenitors.
The presence of primordial binaries would make the BH core-driven balanced evolution (and hence $t_{\rm cc}$) less defined since they would be an ambient source of central energy generation from the beginning of the cluster's evolution. After the segregation of BHs, the central energy generation will be shared with the primordial binaries which could potentially affect the relevant encounter rates among the BHs. }
 
\section*{Acknowledgements}
FA acknowledges support from a Rutherford fellowship 
 (ST/P00492X/1) from the Science and Technology Facilities Council.
 MG acknowledges financial support from the
European Research Council (ERC StG-335936, CLUSTERS).
We thank the anonymous referee for constructive comments that helped to improve an earlier version of 
this paper. 
We thank Carl Rodriguez for sharing the data from 
the Monte Carlo models used in Section~\ref{MCcompS} and for useful 
discussions. We thank Adrian Hamers, Giacomo Fragione, Hagai Perets, and Johan Samsing for
useful discussions.

\bibliographystyle{mnras}

\end{document}